\documentclass[10pt,journal]{IEEEtran}
\usepackage{url}
\usepackage{booktabs} 
\usepackage[noadjust]{cite}
\usepackage{graphicx}
\usepackage[subrefformat=parens, labelformat=parens]{subfig}
\usepackage[font=footnotesize, labelfont=footnotesize]{caption}
\usepackage{caption, float}
\usepackage{float}
\usepackage{amsmath,amssymb,amsfonts}
\usepackage{color}
\usepackage{algorithm, algorithmicx, algpseudocode}
\usepackage{bm, bbold}
\usepackage{tabularx}
\usepackage{enumitem}
\usepackage[dvipsnames]{xcolor}

\newcommand{\memoMK}[1]{{\color{teal} \bf$<$MK: #1$>$}}
\newcommand\addedMK[1]{{\color{black}#1}}
\newcommand\addedCC[1]{{\color{black}#1}}

\begin{document}


\sloppy

\title{Channel and Spectrum Consumption Models for Urban Outdoor-to-Outdoor {28}\thinspace{GHz} Wireless}

\author{Manav Kohli, Carlos E. Caicedo, Tingjun Chen, Irfan Tamim, Angel D. Estigarribia, Tianyi Dai, Igor Kadota, Dmitry Chizhik, Jinfeng Du, Rodolfo Feick, Reinaldo A. Valenzuela, Gil Zussman
\vspace{-\baselineskip}
\thanks{This work was supported by CS3 under NSF grant EEC-2133516; federal and industry funds under the NSF RINGS program, NSF grants CNS-2148128, AST-2232455, AST-2232456, and DGE-2036197; and CONICYT Project FB 0821. A partial and preliminary version of the dataset was presented in~\cite{chen201928}.}
\thanks{M. Kohli was with the Department of Electrical Engineering, Columbia University, New York, NY, 10027, USA. He is now with Apple Inc., Sunnyvale, CA 94086 USA (email: mpk2138@columbia.edu).}
\thanks{C. E. Caicedo is with the School of Information Studies, Syracuse University, Syracuse, NY 13210 USA.}
\thanks{T. Chen is with the Department of Electrical and Computer Engineering, Duke University, Durham, NC 27708 USA.}
\thanks{I. Tamim and G. Zussman are with the Department of Electrical Engineering, Columbia University, New York, NY 10027 USA.}
\thanks{A. D. Estigarribia was with the Department of Electrical Engineering, Columbia University, New York, NY 10027 USA. He is now with Qualcomm, San Diego, CA 92121 USA.}
\thanks{T. Dai was with the Department of Electrical Engineering, Columbia University, New York, NY 10027, USA. He is now with Google, Sunnyvale, CA 94089 USA.}
\thanks{I. Kadota is with the Department of Electrical and Computer Engineering, Northwestern University, Evanston, IL 60208 USA.}
\thanks{D. Chizhik, J. Du, and R. A. Valenzuela are with Nokia Bell Labs, Murray Hill, NJ 07974 USA.}
\thanks{Rodolfo Feick is with the Department of Electronics Engineering, Universidad T\'ecnica Federico Santa Mar\'ia, Valpara\'iso 234000 Chile.}
}



\newcommand{\Intersection}{Int}
\newcommand{\Bridge}{Bri}
\newcommand{\Balcony}{Bal}

\newcommand{\pathgain}{\textsf{PG}}
\newcommand{\snr}{\textsf{SNR}}



\maketitle

\begin{abstract}
\addedMK{Millimeter-wave (mmWave) communication has been widely accepted as an enabler of 6G and other next-generation wireless networks. While mmWave promises extreme data rates and widely available spectrum, high path loss strains link budgets. As it stands, the difficult channel conditions have limited the deployment of mmWave within the 5G NR radio access network (RAN) primarily to dense urban environments. In this paper, we seek to demystify aspects of RAN planning and design for these environments by providing a set of empirical models of the mmWave channel at 28\thinspace{GHz}, alongside a methodology to develop spectrum consumption models (SCMs), which illustrate constraints on spectrum allocation by the RAN.}
\addedMK{We first report on an extensive 28\thinspace{GHz} measurement campaign within the deployment area of the PAWR COSMOS testbed in New York City. This campaign resulted in over 46 million power measurements, collected from over 3,000 links across 24 street sidewalks at four different sites.}
\addedMK{Using these measurements, we study the effects of the setup and environments, such as TX height and seasonal effects. We then derive a series of channel models for path loss and the azimuth beamforming gain loss, and use them to derive distributions of the link SNR values achievable by UEs on the measured sidewalks. We show, among other results, that 100\% of UEs on a given city block can achieve 10\thinspace{dB} SNR at locations with a strong street canyon effect.} \addedCC{Finally, we develop a process to generate SCMs based on the IEEE 1900.5.2 standard using the empirical channel models. The generated SCMs facilitate the evaluation of spectrum sharing and interference management scenarios since they capture all  directional propagation effects reflected in the measurements and provide a way to easily share the main propagation characterization results derived from the measurements.}
\addedMK{We believe that the models, methods, and results in this paper will help inform the future of mmWave wireless network deployments within dense urban areas.}
\end{abstract}


\setlength{\abovedisplayskip}{3pt}
\setlength{\belowdisplayskip}{3pt}

\section{Introduction}
\label{sec:intro}
\addedMK{The widely available spectrum at millimeter-wave (mmWave) frequencies provides opportunity for significantly increased data rates, and will continue to be a key technology for 6G-and-beyond networks}~\cite{hong2021roleofmmw, rappaport2013millimeter, pi2011introduction}. It is also beneficial to higher layer protocols~\cite{ghasempour2018beamtraining} and has the potential to support a broad class of new applications such as high-resolution localization and sensing~\cite{chen2017pseudo}, and augmented/virtual reality~\cite{abari2017enabling}.

To support the design and deployment of mmWave networks, it is important to understand the fundamental propagation properties of mmWave signals and the corresponding effects on the system-level performance in different network scenarios. \addedMK{Spectrum management is a critical factor for network performance, and the ubiquitous use of highly directive antenna systems in mmWave networks presents both a complication and a degree of freedom for the implementation of efficient spectrum management algorithms. Beamforming, made possible by directive antennas~\cite{rappaport2012broadband, rappaport2015wideband, zhang2016openmili}, including phased arrays~\cite{zhang2016openmili, sadhu201728, saha2019x60, bas2018real}, helps to mitigate the high path loss at mmWave frequencies, and therefore must be considered in spectrum management.}
%

%

\begin{figure}[!t]
\centering
\includegraphics[width=\columnwidth]{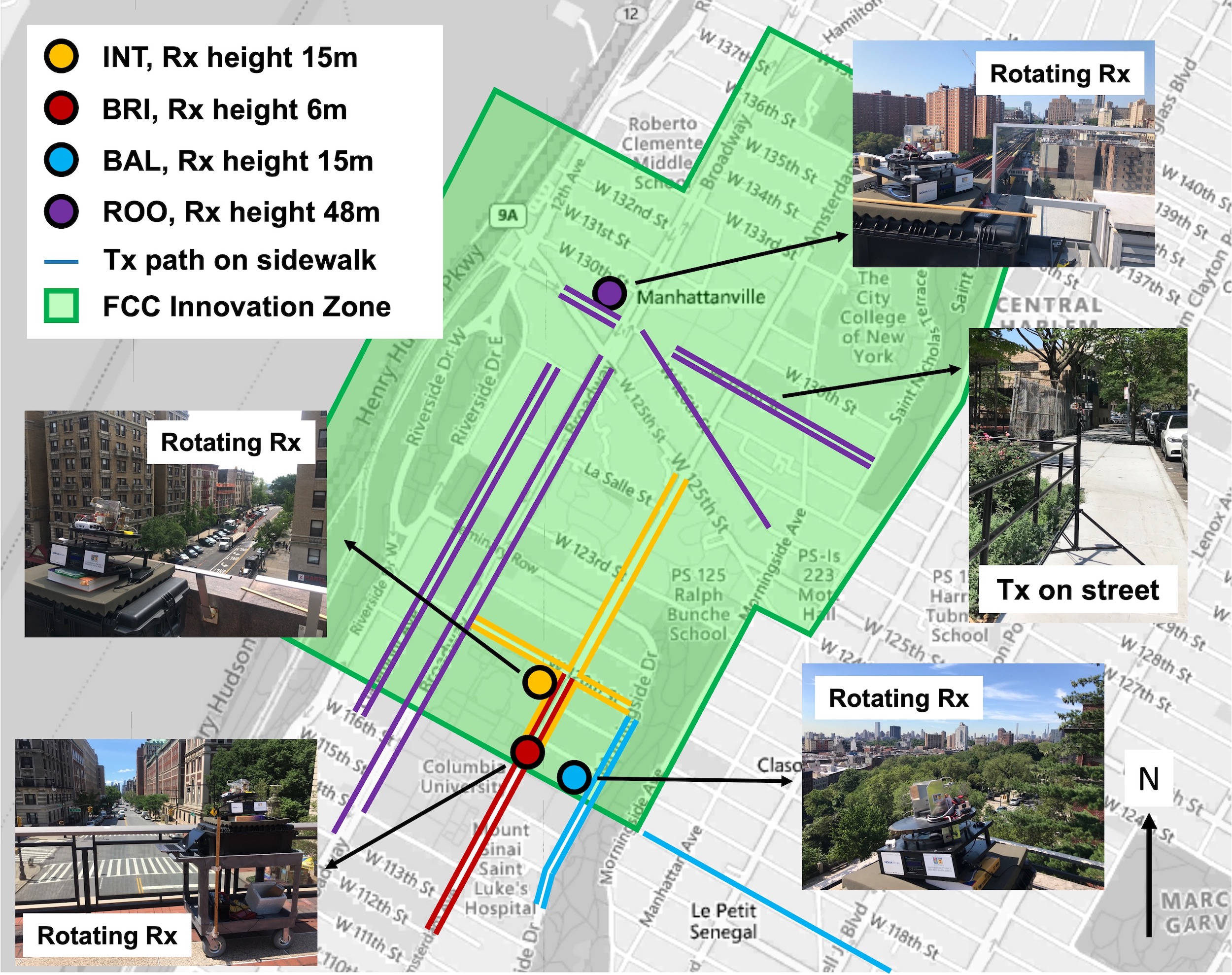}
\vspace{-1\baselineskip}
\caption{Locations and receiver (RX) locations of the four measurement sites within the COSMOS FCC Innovation Zone located in West Harlem, NYC. The RX \addedMK{is placed at one or two fixed locations at each site} (marked by circles), and the transmitter (TX) is moved along sidewalks (marked by lines).}
\label{fig:meas-map}
\vspace{-\baselineskip}
\end{figure}

\addedMK{In this paper, following a discussion of related work and an overview of SCMs, we present several main contributions. First, we present an outdoor-to-outdoor (O--O) measurement campaign conducted within the PAWR COSMOS FCC Innovation Zone~\cite{cosmos, cosmoswebsite, fcc2021innovation}, located within a dense urban environment in West Harlem, New York City. Secondly, we report extensive {28}\thinspace{GHz} channel measurements in this area from four measurement sites as shown in Figure~\ref{fig:meas-map}. Third, we use the channel measurements to investigate various environmental effects, and compute various metrics such as the signal-to-noise ratio (SNR) and corresponding channel capacity in select locations. Lastly, we use the channel measurements in the generation of mmWave spectrum consumption models (SCMs) as standardized by IEEE 1900.5.2~\cite{ scmstdofficial, scmstandard}.}


\noindent\textbf{Measurements: }\addedMK{Our measurement campaign includes over 3,000 link measurements (with over 46 million individual power measurements) collected over 24 scenarios (each corresponding to a street sidewalk) adjacent to the four measurement sites as shown in Figure~\ref{fig:meas-map}. These measurement sites include a fourth-floor balcony overlooking a street intersection, a bridge above a four-lane street, a balcony overlooking a city park, and a tall rooftop next to a raised train track.} \addedMK{These four locations were chosen to best represent the dense, varied urban environment around the COSMOS FCC Innovation Zone. The measurements are collected with a portable 28\thinspace{GHz} narrowband channel sounder, with the transmitter (TX) emulating a UE and a rotating receiver (RX) emulating a 5G NR gNB, together emulating a typical 5G NR RAN link.}

\noindent\textbf{Propagation Models: }\addedMK{We develop path gain models for each O--O scenario using a single-slope exponent and intercept fit to the measured data as a function of link distance, and record the cumulative distribution function (CDF) of the measured azimuth beamforming gain (ABG). We then group the scenarios in two ways: (i) by the four locations and (ii) by \textit{visual} line-of-sight (VLOS) or \textit{visual} non-LOS (VNLOS). With these grouped models, we show a large variation in loss above free space. For example, the \textbf{BAL} location suffers 20\thinspace{dB} of excess loss, whereas \textbf{ROO} only experiences 10\thinspace{dB} of loss above free space. On average, the measured path gain is 5--10\thinspace{dB} higher than that provided by the 3GPP 38.901 urban canyon non-line-of-sight (NLOS) model~\cite{3gpp_los_nlos}. Moreover, the median ABG is within {1.7}\thinspace{dB} and {2.9}\thinspace{dB} of the the nominal value for VLOS and VNLOS sidewalks, respectively.}

\noindent\textbf{Specific Effects: }\addedMK{We also generate comparative models to explore certain effects: (i) swapped TX and RX locations, (ii) raised TX heights, (iii) seasonal effects, and (iv) different placements of the TX on the street or sidewalk. We experimentally observe that most effects are minimal on the considered sidewalks. For example, the path gain fitted lines for TX heights of {1.5}\thinspace{m} and {3}\thinspace{m} differ by only less than {3}\thinspace{dB} over link distances of 40--500\thinspace{m}. Seasonal effects are more significant depending on the sidewalk.}

\noindent\textbf{Channel and Coverage Metrics: }Based on the measurement results, we consider the link SNR values that can be achieved \addedMK{for individual scenarios and aggregated over the four measured locations. In particular, we show that QPSK FR2 data transmission would be supported by more than half of the scenarios beyond 500\thinspace{m} distance between the user equipment (UE) and base station (BS). We also show that the presence of a BS at both ends of a street sidewalk can greatly improve coverage for UEs randomly located along the sidewalks around each BS location, with 100\% of UEs at the \textbf{INT} and \textbf{BRI} locations achieving $>10$\thinspace{dB} 10\textsuperscript{th} percentile SNR 90\%.}

\noindent\textbf{Spectrum Consumption Models: }\addedCC{Finally, we leveraged the collected measurements to generate Spectrum Consumption Models (SCMs) as standardized by the IEEE 1900.5.2 standard~\cite{scmstdofficial, scmstandard, scmwcnc2023} which facilitate the evaluation of spectrum sharing and interference management scenarios since they capture all the directional propagation effects reflected in the measurements and also the directional characteristics of the antennas we used. SCMs can be used in the design of mmWave spectrum management systems and/or the planning of non-interfering mmWave device operations in the urban street canyon environments covered in our measurement scenarios.} 

We believe that our measurement results can provide insights into the deployment of the 28\thinspace{GHz} phased array antenna modules (PAAMs) (developed by IBM and Ericsson~\cite{sadhu201728}), in the COSMOS testbed. The results can also serve as the benchmark for the experimentation with these modules once deployed \addedCC{and the SCMs produced will facilitate the sharing of our measurement results and their use in simulators and spectrum management system design.} 



\section{Related Work}
\label{sec:related}
Various channel measurement campaigns have been conducted for different mmWave frequencies \addedMK{between 28\thinspace{GHz} and 140\thinspace{GHz}} in urban~\cite{rappaport2013millimeter, samimi2013angle, maccartney2014omnidirectional, rappaport2015wideband, molisch2016millimeter, ko2017millimeter, xing2021millimeter, ju2021millimeter, ozdemir202028, kohli2022outdoor} and suburban~\cite{rappaport2012broadband, rappaport2013millimeter, bas201728, zhang201828, du202128, du2020suburban, koymen2015indoor} environments. 

Both directional horn antennas on mechanical steering platforms~\cite{rappaport2012broadband, samimi2013angle, zhang2016openmili, xing2021millimeter, ko2017millimeter, du2020suburban, kohli2022outdoor, koymen2015indoor}, and phased arrays applying beam steering~\cite{zhang2016openmili, sadhu201728, saha2019x60, bas2018real, shkel2021configurable} were considered. \addedMK{Many of these measurement campaigns use wideband} channel sounders, \addedMK{which} can provide measurements of not only the path gain, but also other metrics such as the power delay profile. However, most of these campaigns consider a limited number of measurement links (usually at the order of 10s \addedMK{or 100s}) with different Tx and Rx locations. Recent work also demonstrates link-level channel measurements using commercial 802.11ad devices~\cite{sur2015flexible, steinmetzer2017compressive, shkel2021configurable}.

\addedMK{In contrast to many prior measurement efforts,} we use a narrowband channel sounder that, can uniquely support (i) fast recording of received signal on the entire 360$^{\circ}$ azimuth plane (allowing us to obtain the effective azimuth beamforming gain), and (ii) easy and continuous measurements on relatively long streets with small measurement separation distances (see Section~\ref{sec:meas}). Hence, our measurement campaign focuses on accurately characterizing the mmWave channel \emph{along individual streets} at many link distance values. This results in an extensive amount of collected measurement data.

\begin{figure*}[!t]
\centering
\subfloat[\textbf{INT}]{
\includegraphics[width=0.23\linewidth]{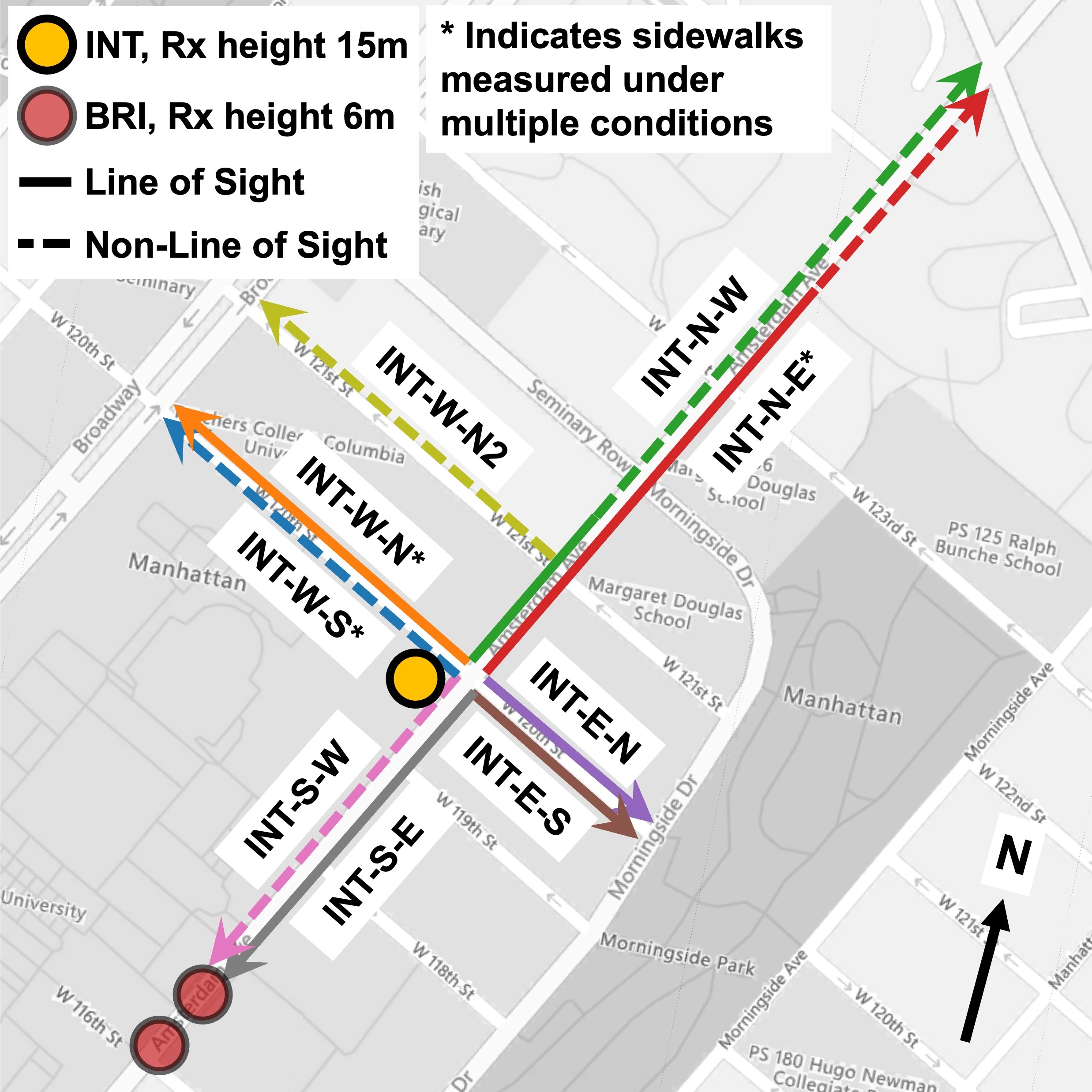}
\label{fig:loc-int}
}
\hspace{1pt}
\subfloat[\textbf{BRI}]{
\includegraphics[width=0.23\linewidth]{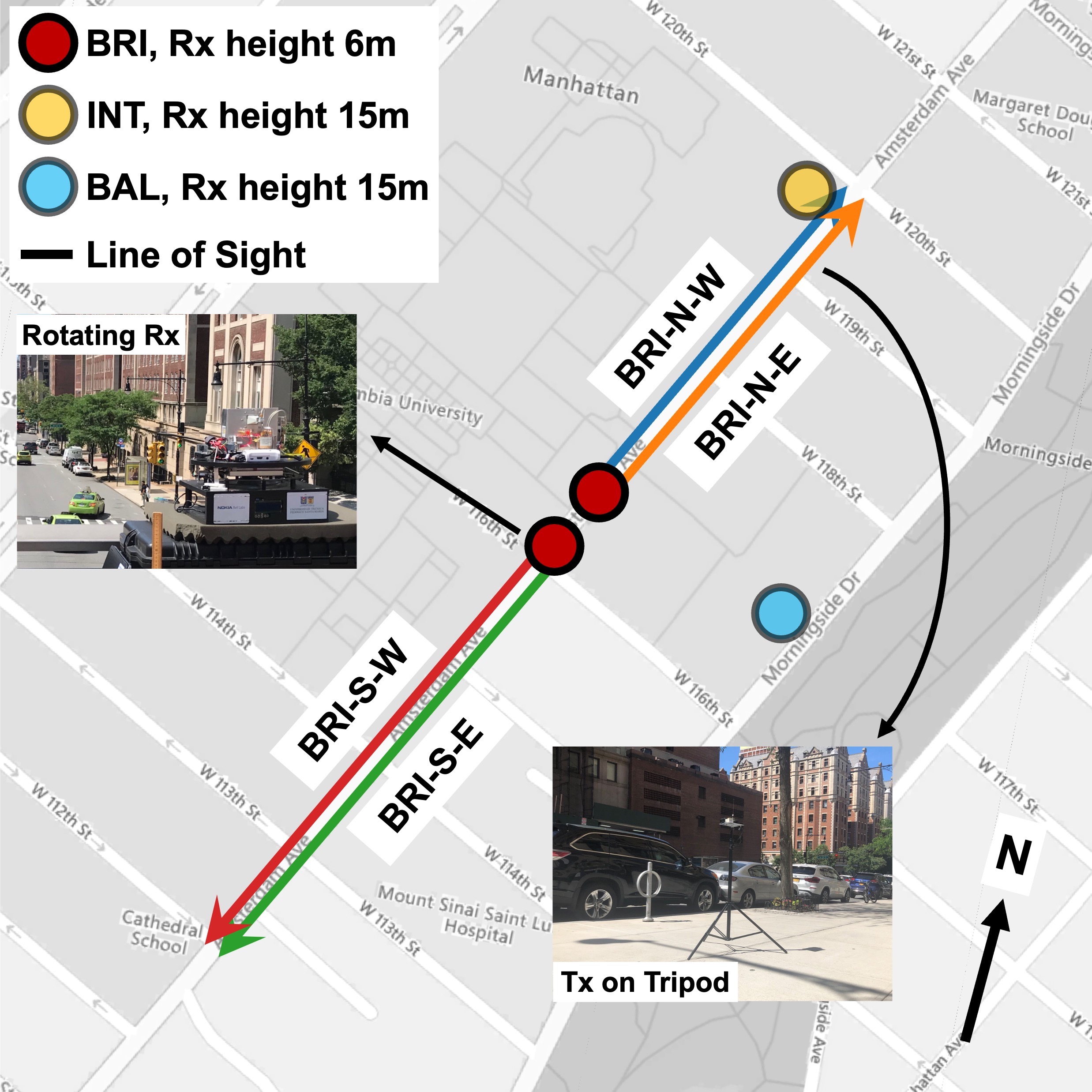}
\label{fig:loc-bri}
}
\hspace{1pt}
\subfloat[\textbf{BAL}]{
\includegraphics[width=0.23\linewidth]{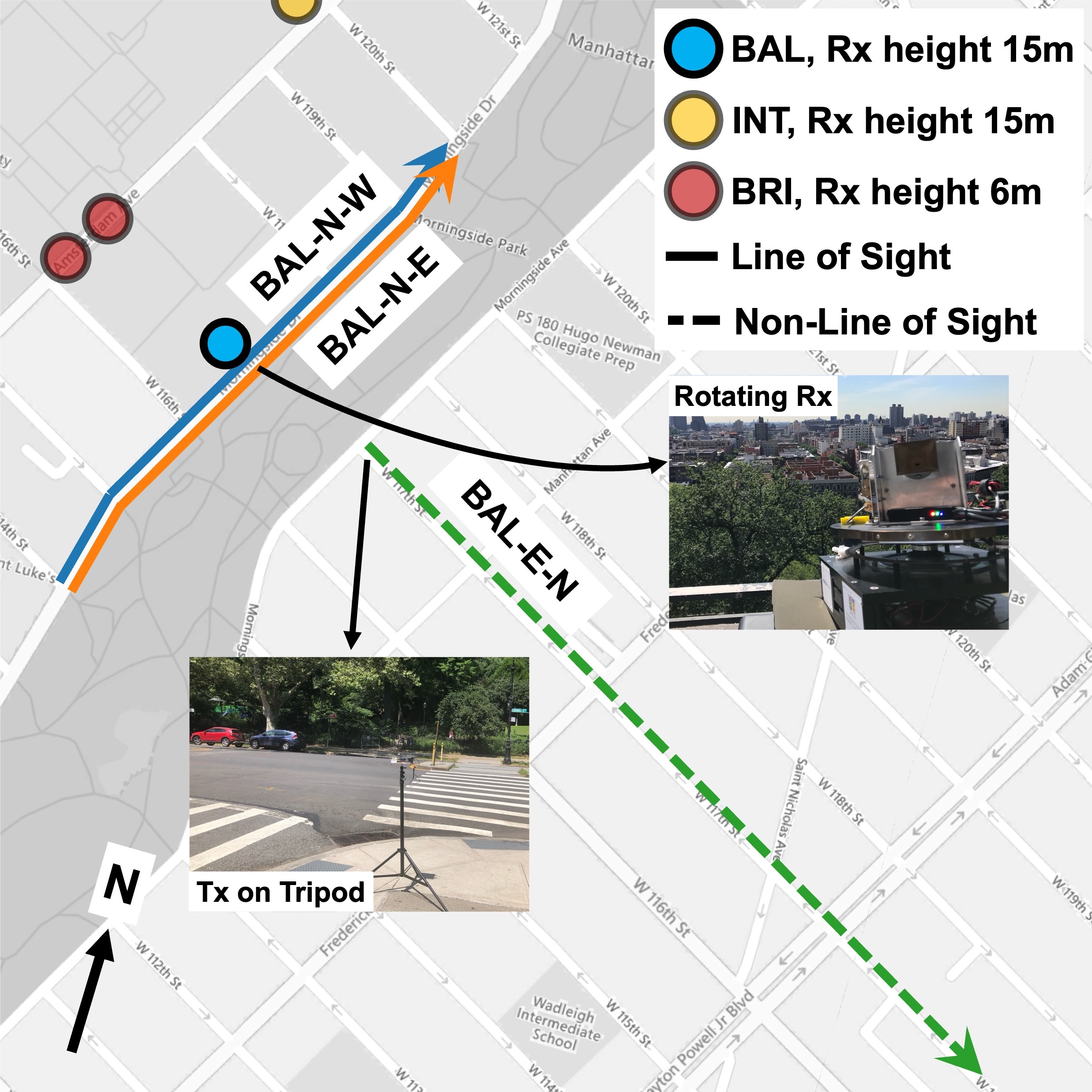}
\label{fig:loc-bal}
}
\hspace{1pt}
\subfloat[\textbf{ROO}]{
\includegraphics[width=0.23\linewidth]{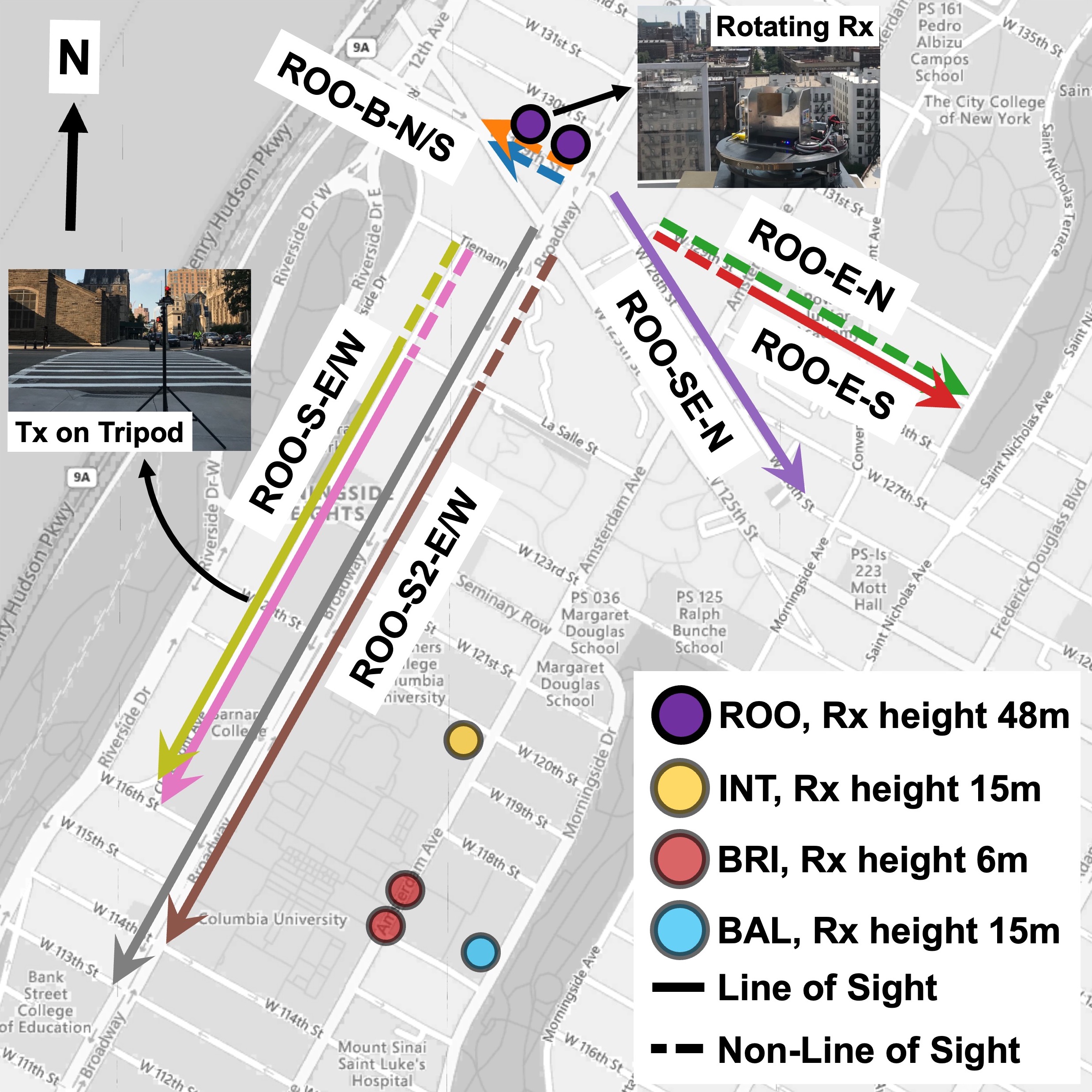}
\label{fig:loc-roo}
}
\caption{\memoMK{Need to update LOS / NLOS to VLOS / NVLOS}\addedMK{Maps of sidewalk measurements at each of the four RX locations.}}
\label{fig:locmaps}
\vspace{-\baselineskip}
\end{figure*}

 \addedMK{Prior work on spectrum management for mmWave networks has made use of ITU recommendation P.676-12~\cite{iturp67612, marcus2023millimeter} to model path loss for the goal of informing management decisions. Spectrum management methods have also been considered various forms of mmWave networks, including 5G/6G cellular networks~\cite{saha2021millimeter, teixeira2019spectrum, liu2021resource, hanpinitsak2023maximum} and UAV swarm networks~\cite{feng2019spectrum}. In this work, we use the extensive set of collected measurement data to develop IEEE 1900.5.2 SCMs~\cite{scmstandard}, which may be used in spectrum management methods and algorithms. Therefore, to the best of our knowledge, this work is the first at \textit{presenting a method for developing a standardized SCM-based representation of urban mmWave spectrum usage, based on an extensive set of real-world measurements in a dense urban environment}.}


\section{Overview of Spectrum Consumption Models}
\label{sec:scm}
\addedCC{SCMs provide a mechanism for RF devices to “announce” or “declare” their intention to use the spectrum and their needs in terms of interference protection, and also to determine compatibility (i.e., non-interference) with existing devices. SCMs capture, and allow RF engineers to operate with all the directional dependent information required to manage spectrum (e.g. Antenna radiation pattern, antenna orientation, propagation attenuation, device trajectory) to achieve non-interfering mmWave operational communication environments \cite{scmstandard,scmwcnc2023}. }

\addedCC{SCMs use a set of 11 data elements, referred to as constructs, to describe the spectral, spatial, and temporal characteristics of spectrum use by any RF device and/or system. These constructs, which are defined in the IEEE 1900.5.2 standard~\cite{scmstdofficial,scmstandard}, can be used to build different types of SCMs, including:}
    (i) \addedCC{\textit{\textbf{TX models}} that convey the extent and strength of RF emissions from a TX;}
    (ii) \addedCC{\textit{\textbf{RX models}} that convey what harmful interference to an RF RX device is; and}
    (iii) \addedCC{\textit{\textbf{System and Set models}} that group several TX and RX SCMs.}

\addedCC{The SCM constructs most relevant to this work are described below.} 
\addedCC{Unless otherwise stated, the constructs must be used in both TX and RX models.}
\begin{itemize}[leftmargin=*]
    \addedCC{\item \textit{\textbf{Reference Power:}} Value used as the reference power level for the emission of a TX or for the allowed interference in a RX. It is used in the spectrum mask and underlay mask constructs.}
    \addedCC{\item \textit{\textbf{Spectrum mask:}} Defines the relative spectral power density of emissions by frequency. This construct is mandatory for Transmitter models only.}
    \addedCC{\item \textit{\textbf{Underlay mask:}} Defines the relative spectral power density of allowed interference by frequency. This construct is mandatory for Receiver models only.}
    \addedCC{\item \textit{\textbf{Power map:}} Defines a relative power flux density per solid angle. It conveys the dispersion of electromagnetic energy from a TX antenna or the concentration of energy at a RX antenna.}
    \addedCC{\item \textit{\textbf{Propagation map:}} Defines a path loss per solid angle (direction).}
    \addedCC{\item \textit{\textbf{Schedule:}} Specifies the time in which the model applies (start time, end time). Periodic activity details can also be defined.}
    \addedCC{\item \textit{\textbf{Location:}} Specifies where an RF device may be used. Several types of locations are supported: a point, a volume, a trajectory or an orbit.}
\end{itemize}

\addedCC{Additional SCM constructs can be used to describe intermodulation effects, policy or protocol based spectrum coexistence, and broadcasting behavior. Those constructs are: Intermodulation mask, Platform name, Minimum power spectral flux density, and Policy/Protocol.

Using the information provided by the SCMs, any entity can use the methods standardized in IEEE 1900.5.2 to evaluate whether the spectrum use of two or more systems is compatible. The compatibility computation indicates whether the systems can coexist without causing harmful interference to each other or if they will interfere with each other (i.e., they are non-compatible). This computation can be extended to scenarios with multiple devices where aggregate interference effects need to be taken into account to make a final compatibility decision. }


\section{Equipment, \addedMK{Locations}, and Dataset}
\label{sec:meas}
\addedMK{In this section, we provide details on the equipment and locations used during the measurement campaign, as well as an overview of the resulting dataset.}

\subsection{Measurement Equipment}
\label{ssec:meas-equipment}
\addedMK{We utilize a narrowband 28\thinspace{GHz} channel sounder consisting of a separate TX and RX, shown in the inset photographs within Figs.~\ref{fig:meas-map} and~\ref{fig:locmaps}. The TX is equipped with an omnidirectional antenna with 0\thinspace{dBi} azimuthal gain, and transmits a +22\thinspace{dBm} continuous-wave tone. The RX, mounted on a rotating platform, is fed by a horn antenna with 14.5\thinspace{dBi} gain and 10$^{\circ}$ half-power beamwidth in azimuth and 24\thinspace{dBi} gain in total.}
\addedMK{Both the TX and RX have free running local oscillators (LOs) with a frequency offset of at most 10\thinspace{kHz}, and are powered by batteries to support portable measurements.}
\begin{figure}[t]
\centering
\subfloat[]{
\includegraphics[height=1.7in]{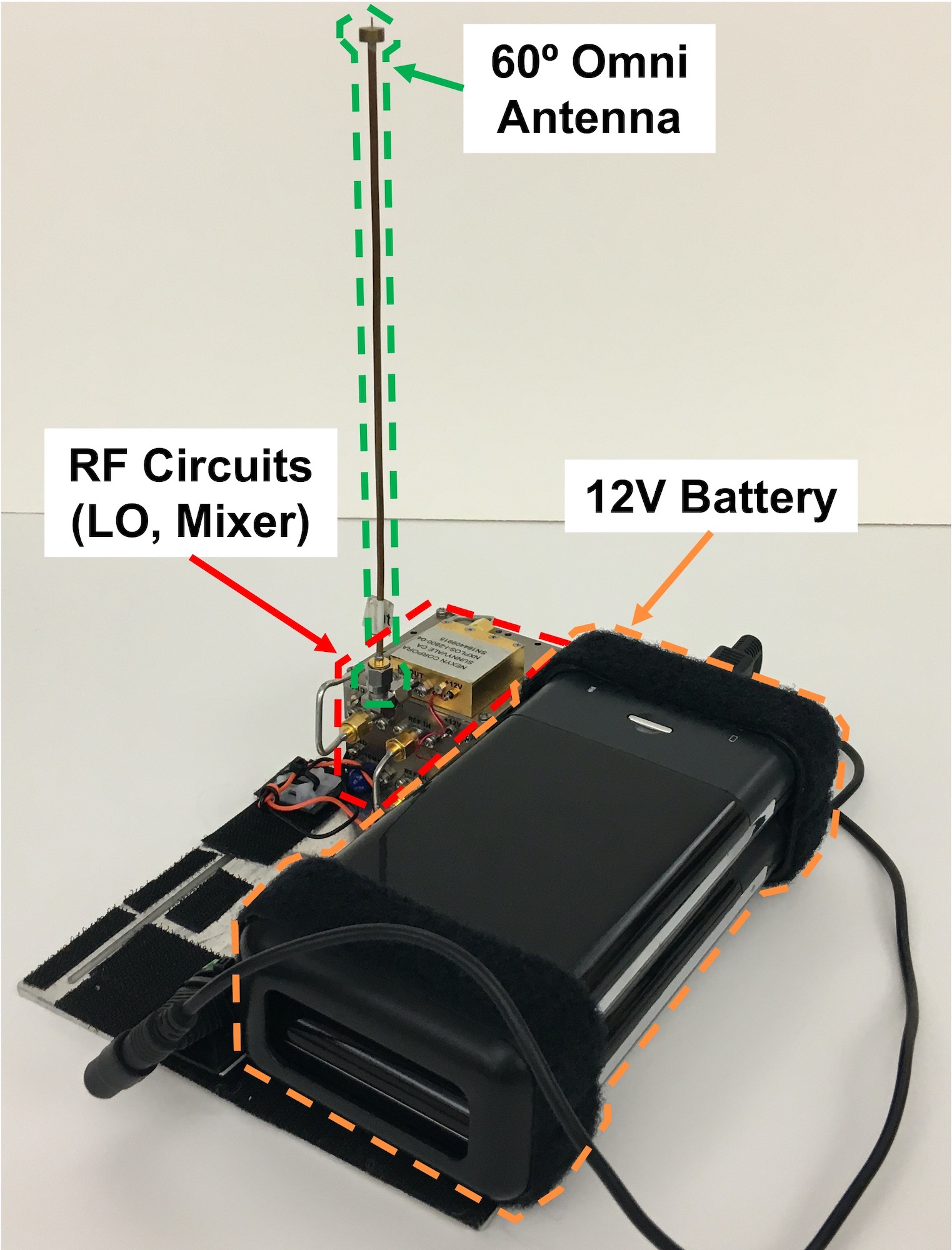}
\label{fig:tx-fig}
}
\hspace{2pt}
\subfloat[]{
\includegraphics[height=1.7in]{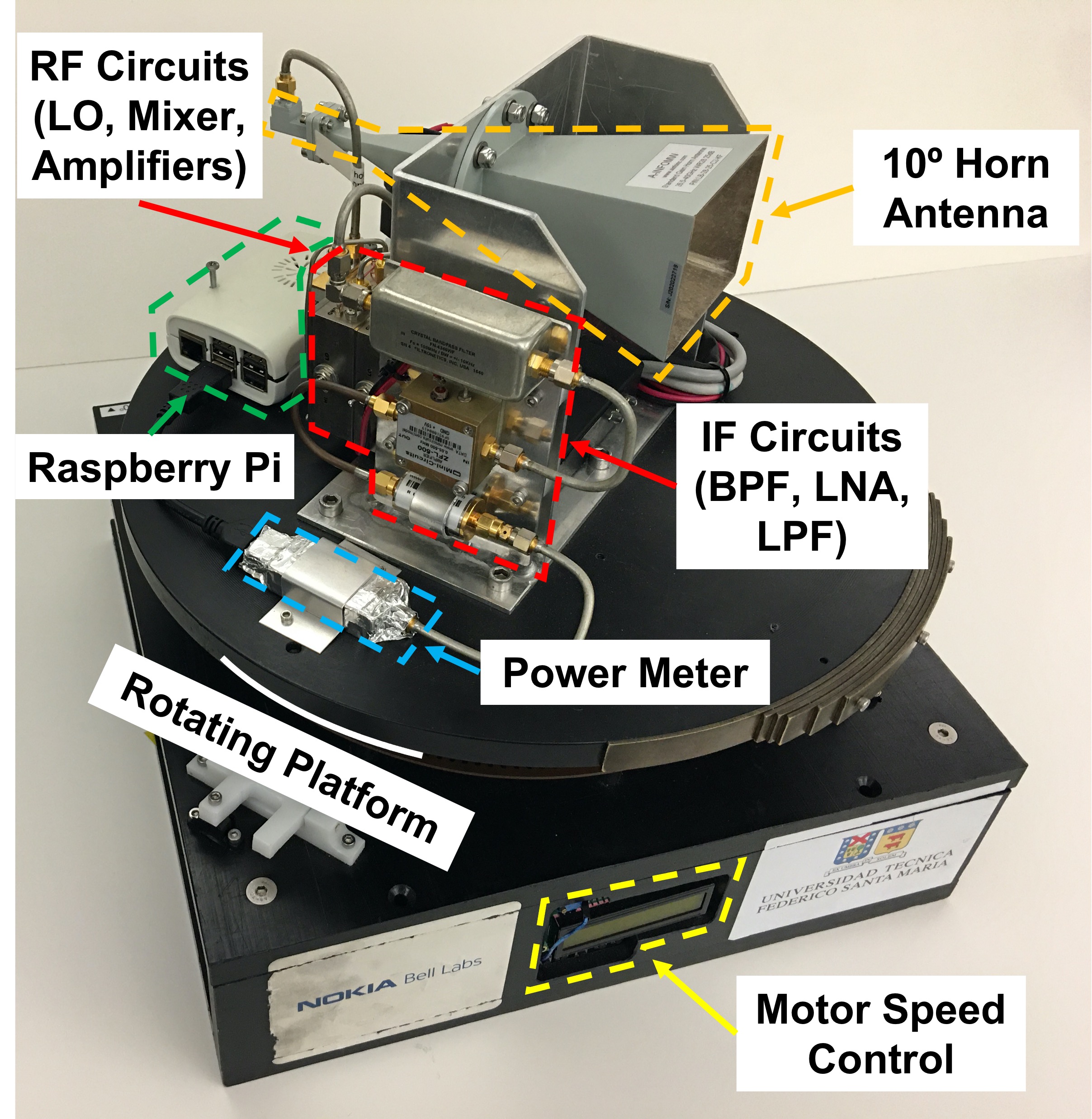}
\label{fig:rx-fig}
}
\caption{\addedMK{Labelled photographs of the custom 28\thinspace{GHz} (a) TX and (b) RX.}}
\label{fig:txrx-fig}
\vspace{-\baselineskip}
\end{figure}

The channel sounder is calibrated within an anechoic chamber to ensure an absolute power accuracy of {0.15}\thinspace{dB}. \addedMK{In the configuration with the rotating horn antenna, the channel sounder can measure path gains between -62\thinspace{dB} and -161\thinspace{dB} with at least 10\thinspace{dB} of SNR to provide a reliable measurement in the presence of channel fading. This corresponds to a {1,000}\thinspace{m} TX-RX link distance with {40}\thinspace{dB} of additional loss above free space. In practice, we are able to achieve sufficient SNR for reliable path loss measurement at distances well in excess of {1,000}\thinspace{m}, as is the case for some scenarios at  the \textbf{ROO} location. Additional details on the channel sounder are provided in~\cite{du2020suburban,chen201928,kohli2024outdoor}.}

\subsection{Measurement \addedMK{Locations}}
\label{ssec:meas-environment}

We perform extensive {28}\thinspace{GHz} channel measurements in the deployment area of the PAWR COSMOS testbed~\cite{cosmos,yu2019cosmos}, which is currently being deployed in West Harlem, New York City. \addedMK{Fig.~\ref{fig:meas-map} shows a map encompassing the entire measurement area alongside example street views of the measurement locations. Detailed maps of each individual measurement location are provided in Fig.~\ref{fig:locmaps}.} The considered area is a representative \emph{urban street canyon} environment containing mostly concrete and brick buildings with heights of {10--60}\thinspace{m} (3--16 stories). The vegetation in the measurement area consists of sparse, \addedMK{young trees equally spaced} on both sides of the city streets, \addedMK{a typical feature of cities in the northeastern United States. As shown in the maps, the area is also contains two parks with dense vegetation.}

\begin{table*}[t!]
\centering
\caption{\addedMK{Summary of 24 O--O scenarios with computed path gain model and median azimuth beamforming gain.}} 
\label{tab:meas-summary}
\footnotesize
\begin{tabular}{|c|c|c|c|c|c|c|c|c|c|}
\hline
\shortstack{Sidewalk \\ Name} & \shortstack{Sidewalk \\ Length (m)} & \shortstack{Meas. Step \\ Size (m)} & \shortstack{\# Links} & \shortstack{Slope \\ ($n$)} & \shortstack{Intercept \\ ($b$\thinspace[dB])} & \shortstack{RMS \\ ($\sigma$\thinspace[dB])} & \shortstack{Median Azi. \\  Gain (dBi)} & \shortstack{10$^{\textrm{th}}$-perc. Azi. \\ Gain (dBi)} & \shortstack{CW angle\\rel. to N}\\
\hline
INT-N-E & 507 & 3/6 (near/far) & 101 & $-$3.5 & $-$36.8 & 4.3 & 14.1 & 12.3 & 120\\
\hline
INT-W-N & 256 & 3 & 79 & $-$2.6 & $-$52.2 & 4.4 & 14.2 & 13.2 & 120\\
\hline
INT-S-E & 317 & 3 & 93  & $-$3.4 & $-$35.5 & 4.6 & 14.2 & 13.4 & 120\\
\hline
INT-E-N & 146 & 1.5 & 85 & $-$2.3 & $-$60.3 & 4.5 & 14.2 & 13.1 & 120\\
\hline
INT-E-S & 146 & 1.5 & 88 & $-$2.8 & $-$49.5 & 3.2 & 14.1 & 12.4 & 120\\
\hline
INT-N-W & 509 & 3 & 139 & $-$3.6 & $-$36.0 & 3.6 & 13.1 & 11.8 & 120\\
\hline
INT-W-S & 256 & 3 & 69 & $-$3.1 & $-$47.5 & 3.1 & 11.6 & 10.4 & 120\\
\hline
INT-S-W & 317 & 3 & 100 & $-$3.6 & $-$39.2 & 3.4 & 12.9 & 11.2 & 120\\
\hline
BRI-N-E & 219 & 3 & 65 & $-$2.3 & $-$60.0 & 3.9 & 12.6 & 11.3 & 30\\
\hline
BRI-N-W & 219 & 3 & 70 & $-$2.6 & $-$52.5 & 4.3 & 13.4 & 11.7 & 30\\
\hline
BRI-S-E & 280 & 3/6 & 84 & $-$2.5 & $-$55.7 & 5.5 & 12.8 & 11.6 & 210\\
\hline
BRI-S-W & 280 & 3/6 & 87 & $-$2.2 & $-$59.8 & 4.0 & 13.2 & 11.4 & 210\\
\hline
\addedMK{BAL-N-E} & 488 & 3 & 156 & $-$3.4 & $-$47.2 & 5.8 & 13.9 & 12.8 & 208\\
\hline
BAL-N-W & 464 & 3 & 136 & $-$2.9 & $-$66.9 & 4.2 & 12.6 & 10 & 208\\
\hline
\addedMK{BAL-E-N} & 842 & 3/6/9 & 129 & $-$1.5 & $-$94.1 & 6.5 & 14.1 & 12.1 & 120\\
\hline 
ROO-B-N & 98 & 3 & 33 & 0.28 & $-$108.1 & 3.5 & 10.5 & 8.0 & 300\\
\hline
ROO-B-S & 98 & 3  & 33 & 4.45 & $-$179.0 & 3.6 & 10.9 & 8.7 & 300\\
\hline
ROO-S-W & 1058 & 3/6/9 & 150 & $-$1.00 & $-$103.8 & 5.7 & 13.7 & 12.0 & 300\\
\hline
ROO-S-E & 1040 & 3/6/9 & 137 & $-$0.21 & $-$119.7 & 5.9 & 13.8 & 12.7 & 300\\
\hline
\addedMK{ROO-S2-E} & 1102 & 3/6/9/16 & 171 & $-$1.14 & $-$106.2 & $7.9$ & 13.6 & 11.6 & 30 \\
\hline
\addedMK{ROO-S2-W} & 1209 & 3/6/9/16 & 118 & $-$3.05 & $-$47.6 & $4.6$ & 14.1 & 12.4 & 30 \\
\hline
ROO-SE-N & 573 & 3 & 114 & $-$2.36 & $-$62.8 & 5.5 & 14.0 & 12.5 & 120 \\ 
\hline
ROO-E-N & 647 & 3/6 & 97 & $-$2.19 & $-$7.8 & 3.5 & 13.4 & 12.2 & 120\\
\hline
ROO-E-S & 644 & 3/6 & 97 & 0.67 & $-$141.5 & 5.8 & 14.1 & 12.5 & 120\\
\hline
\end{tabular}
\end{table*}

\begin{table*}[t!]
\centering
\caption{\addedMK{Summary of eight additional O--O scenarios taken to study additional effects.}}
\label{tab:meas-summary-addl}
\footnotesize
\begin{tabular}{|c|c|c|c|c|c|c|c|c|c|c|}
\hline
\shortstack{Sidewalk \\ Name} & \shortstack{Condition} & \shortstack{Sidewalk \\ Length (m)} & \shortstack{\# of Meas. \\ Links} & \shortstack{Slope \\ ($n$)} & \shortstack{Intercept \\ ($b$\thinspace[dB])} & \shortstack{RMS \\ ($\sigma$\thinspace[dB])} & \shortstack{Median Azi. \\ Gain (dBi)} & \shortstack{10$^{\textrm{th}}$-perc. Azi. \\ Gain (dBi)} & \shortstack{CW angle\\rel. to N}\\
\hline
INT-N-E-NLe & No tree leaves & 507 & 125 & $-$2.95 & $-$42.9 & 3.8 & 14.1 & 12.4 & 120\\\hline
INT-N-E-10ft & Tx raised to 10\thinspace{ft} & 507 & 79 & $-$3.86 & $-$27.0 & 3.7 & 13.9 & 11.0 & 120\\\hline
INT-W-N-NLe & No tree leaves & 256 & 77 & $-$1.92 & $-$63.9 & 3.7 & 14.1 & 12.5 & 120 \\\hline
INT-W-N-Swap & Tx and Rx swapped & 256 & 79 & $-$2.72 & $-$48.7 & 3.2 & 12.9 & 10.6 & 210 \\\hline
INT-W-N-St & Tx on street & 256 & 68 & $-$2.33 & $-$56.4 & 3.6 & 13.7 & 9.9 & 120 \\\hline
INT-W-N2 & Not adjacent to Rx & 259 & 81 & $-$5.58 & $-$22.1 & 6.3 & 12.9 & 9.8 & 120 \\\hline
INT-W-S-Swap & Tx and Rx swapped & 256 & 79 & $-$2.93 & $-$52.8 & 2.9 & 10.4 & 8.2 & 210 \\\hline
INT-W-S-Wall & Tx next to wall & 198 & 53 & $-$3.32 & $-$40.7 & 3.5 & 11.9 & 9.4 & 120 \\\hline 
\end{tabular}
\vspace{-0.5\baselineskip}
\end{table*}

We consider \addedMK{four measurement sites} where the RX is placed to emulate a \addedMK{5G NR gNB, or colloquially a ``base station'' (BS)},  representing four different deployment scenarios. \addedMK{The TX is moved along sidewalks surrounding the RX location, representing a user equipment (UE). Each of the four locations are shown, respectively, in Fig.~\ref{fig:locmaps}, and described in additional detail in the list below.}
\begin{itemize}[leftmargin=*, topsep=0pt]
\item
\addedMK{Four-way \textbf{int}ersection (\textbf{INT}): a 4$^{\textrm{th}}$ floor outdoors balcony at the southwest} corner of the intersection of Amsterdam Avenue and West 120$^{\textrm{th}}$ street.
\addedMK{The balcony is at a height of 15\thinspace{m} above the surrounding sidewalk.} This location emulates the scenario where the BS is deployed on \addedMK{the rooftop or side of a corner building within} in an urban street canyon, a very common situation in urban areas.
\item
\addedMK{Pedestrian \textbf{bri}dge (\textbf{BRI}): the center of a pedestrian bridge (with a height of {6}\thinspace{m} above the street below)} crossing  Amsterdam Avenue between West 116$^{\textrm{th}}$ and 117$^{\textrm{th}}$ streets. This emulates the scenario where the BS is deployed on a lightpole in the middle of a two-way avenue.
\item
\addedMK{Open \textbf{bal}cony overlooking a city park (\textbf{BAL})}: the balcony of \addedMK{an events venue with a height of {15}\thinspace{m} above the surrounding sidewalks} facing Morningside Park, where the neighboring buildings are 4--16 stories high. This location uniquely does \textit{not} meet the definition of the urban street canyon, as one side of the street does not contain buildings. This emulates a scenario where the BS is deployed on a building roof or side facing an open-space area.
\item
\addedMK{Tall \textbf{roo}ftop (\textbf{ROO}): the rooftop terrace of the Jerome L. Greene Science Center, at the intersection of Broadway and 129$^{\textrm{th}}$ street. This location is at a height of {45-50}\thinspace{m}, which emulates the deployment of a BS located at significant height above the street.}
\end{itemize}

\noindent\addedMK{Each sidewalk measurement is represented with the following notation: \textbf{LOC}-$D$-$S$-$C$, where $D$ is the direction of the street extending away from the RX location, $S$ indicates which of the two sidewalks is being measured, and $C$ is an additional condition used for measurements and described in Table~\ref{tab:meas-summary-addl}. As an example, the sidewalk \textbf{INT}-W-N-NLe denotes the \textbf{n}orthern sidewalk travelling \textbf{w}estwards from the \textbf{INT} RX location, with the condition of \textbf{n}o \textbf{le}aves present on the sidewalk trees.}

Measurements are collected on 8/4/3/9 sidewalks
\addedMK{at \textbf{INT}/\textbf{BRI}/\textbf{BAL}/\textbf{ROO}, respectively, with an additional eight measurements taken on the same sidewalks at \textbf{INT} to evaluate the specific conditions described in Section~\ref{sec:validation}. Each sidewalk measurement represented in Fig.~\ref{fig:meas-map} by a line and in Fig.~\ref{fig:locmaps} by a (dashed) arrow indicating both the direction and whether the sidewalk was in VLOS or VNLOS to the RX. It is typically not possible to conclusively determine whether a given link is truly LOS or NLOS, due to the complex and time-varying urban environment containing foliage, vehicles, and other street clutter. Therefore, we make use of the consistent and empirical VLOS/VNLOS categorization for each TX-RX link, judged during measurement based on whether the RX location may be seen from the TX location on the sidewalk. Note that these definitions are distinct from the definitions of LOS/NLOS that may be used by 3GPP or other standards, and we investigate this in more detail in Section~\ref{sssec:results-pg-allloc}.}



%

\addedMK{At} each site, the rotating RX with a 10$^{\circ}$ horn antenna is used to emulate the BS. \addedMK{The RX rotates at 120 revolutions per minute, providing close to 1$^{\circ}$ angular resolution for capturing} signals arriving from all azimuth directions. The TX with an omni-directional antenna, emulating a \addedMK{UE, is mounted} on a tripod and \addedMK{moved linearly} along the sidewalks to obtain link measurements as a function of the \addedMK{three-dimension Euclidean} link distance $d$. On most sidewalks, the TX is moved away from the RX at a step size of {3}\thinspace{m} for short distances, and {6}\thinspace{m} or {9}\thinspace{m} for long distances (i.e., $d > $ {250}\thinspace{m}). \addedMK{For reference, a regular New York City block spans roughly 80\thinspace{m} north-south and 250\thinspace{m} east-west.}

\addedMK{We found that the sidewalk paving tiles are a highly uniform {1.5}\thinspace{m} $\times$ {1.5}\thinspace{m} within the measurement area, and therefore provide an accurate reference for measuring distance from the RX along the sidewalk.} Some exceptional sidewalks are (i) \textbf{INT}-E-N/S, where the TX is moved away from the RX at a smaller step size of {1.5}\thinspace{m} to ensure a sufficient number of measurement links\addedMK{, and (ii) \textbf{ROO}-S2-E/W, where the TX was moved at step sizes of {16}\thinspace{m} at distances beyond 800\thinspace{m}.}

For a given pair of TX and RX locations, we record one link measurement which lasts for {20}\thinspace{s} and consists of \addedMK{at least} 40 full 360$^{\circ}$ azimuth scans. The link distance of each measurement is calculated using the 3D geometric coordinates obtained from Google Earth with terrain characteristics (e.g., the elevation/slope of each sidewalk) taken into account. For example, sidewalks Int-N-W and Int-N-E are going downhill from south to north, and the resulting relative height between the RX and TX is {15/41}\thinspace{m} at the nearest/furthest (southernmost/northernmost) location, respectively. 

Table~\ref{tab:meas-summary} summarizes our measurement campaign on the 24 sidewalks between March 2019 and January 2021, where the sidewalk color corresponds to the colored lines in Fig.~\ref{fig:meas-map}. \addedMK{Table~\ref{tab:meas-summary-addl} lists eight additional measurements collected to study specific effects described in Section~\ref{sec:validation}. Over 3,000 link measurements were collected. For each link measurement, at least 16,000 power samples are collected as a function of both time and azimuth angle, with the rotating RX spinning for at least {20}\thinspace{seconds}. From these power measurements,} the path gain and effective azimuth gain can be computed (see Section~\ref{ssec:meas-metrics}). \addedMK{Overall, over 41 million power measurements were collected across the 32 measurement scenarios.}

\begin{figure*}[t]
\vspace{-1\baselineskip}
\subfloat[]{
\includegraphics[width=0.3\linewidth]{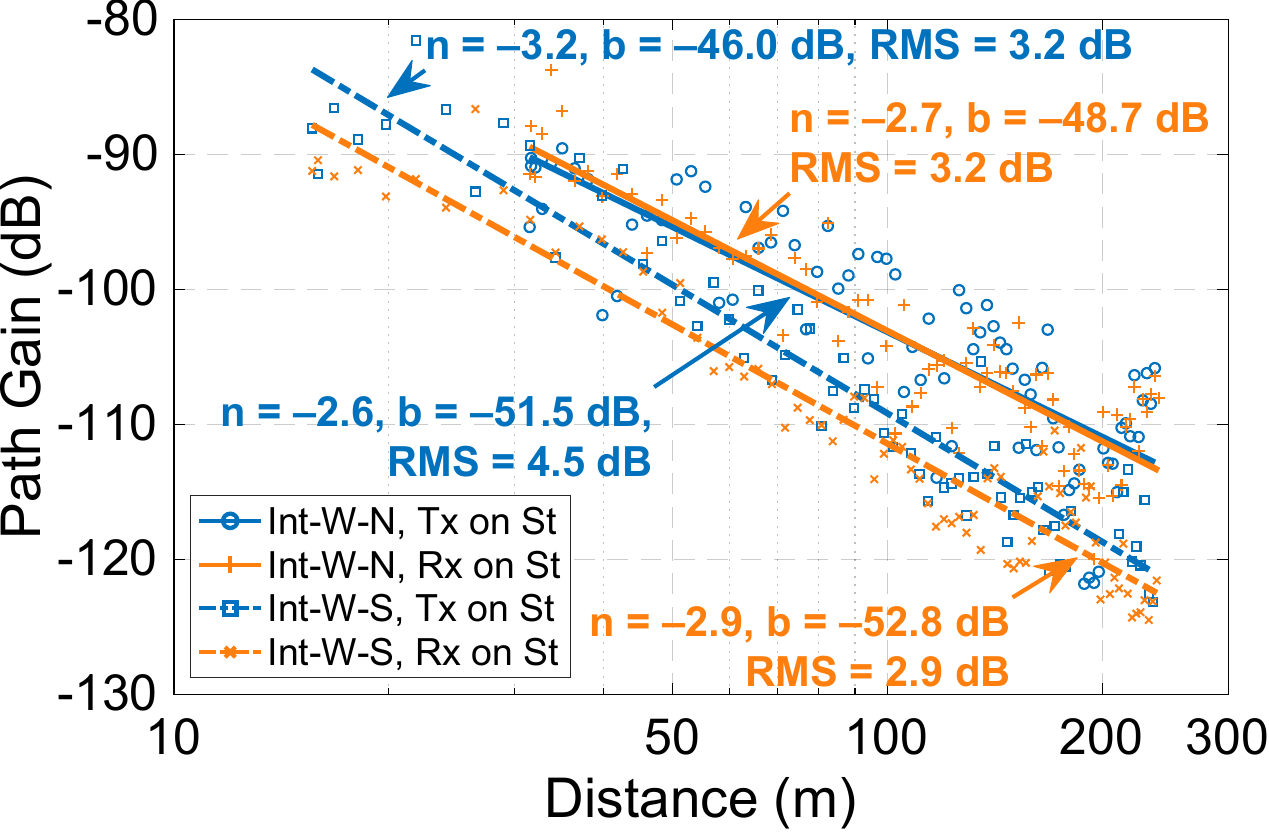}
\vspace{-1\baselineskip}
\label{fig:txrxswap-compare-pg}}
\hspace{10pt}
\setcounter{subfigure}{2}
\subfloat[]{
\includegraphics[width=0.3\linewidth]{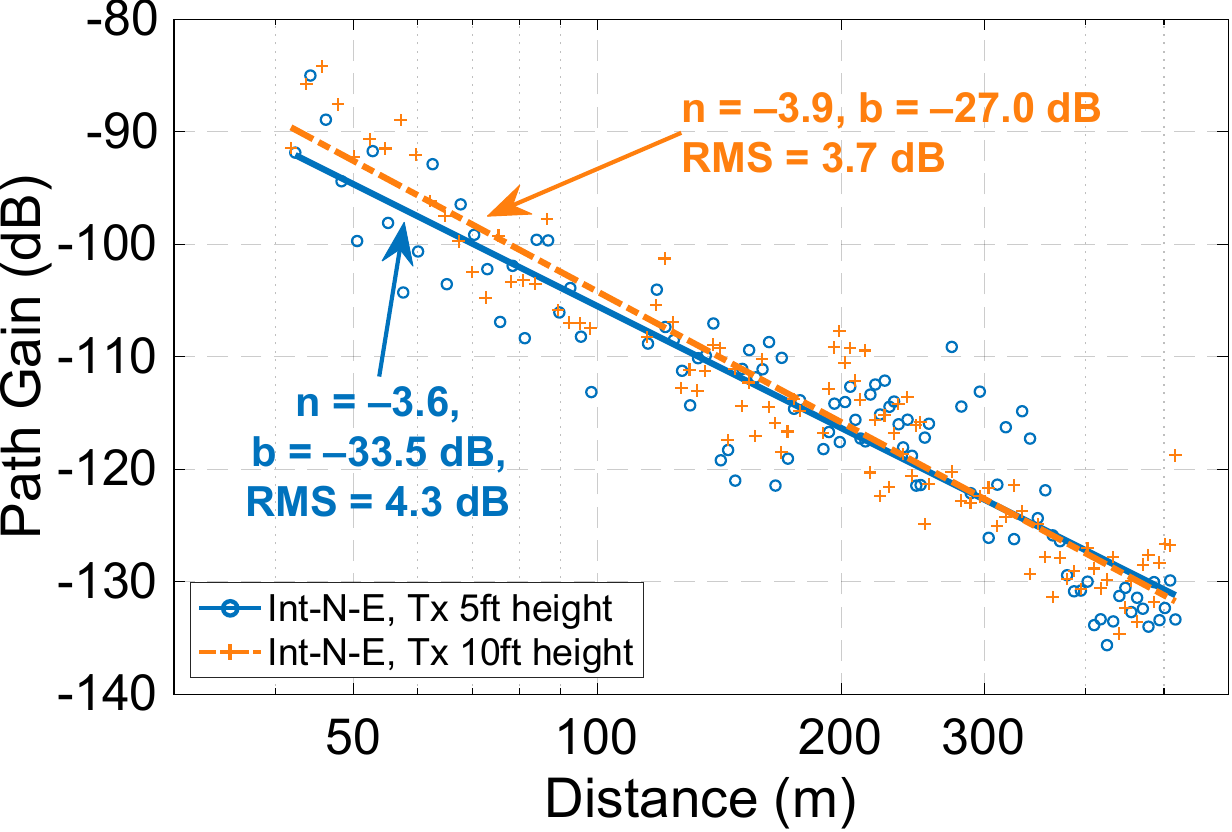}
\vspace{-1\baselineskip}
\label{fig:txheight-compare-pg}}
\hspace{10pt}
\setcounter{subfigure}{4}
\subfloat[]{
\includegraphics[width=0.3\linewidth]{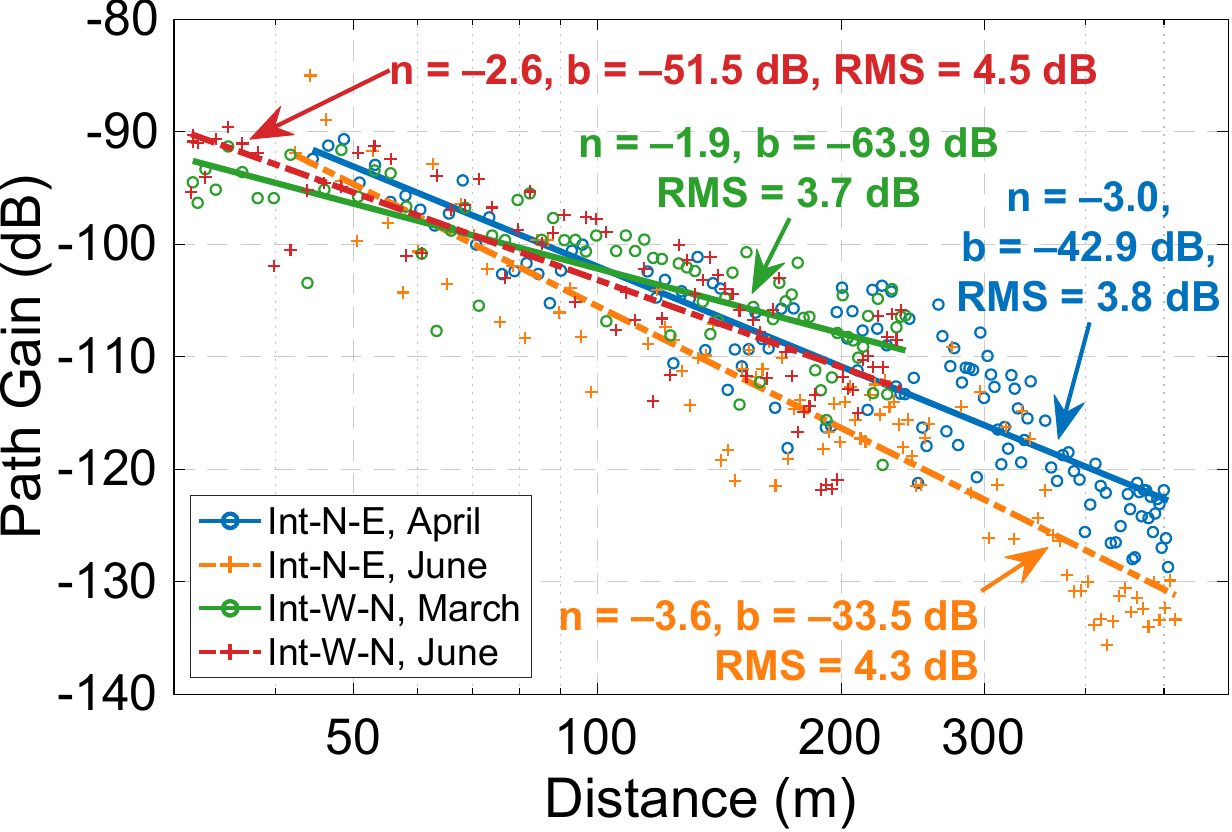}
\vspace{-1\baselineskip}
\label{fig:seasonal-compare-pg}}
\setcounter{subfigure}{1}
\subfloat[]{
\includegraphics[width=0.3\linewidth]{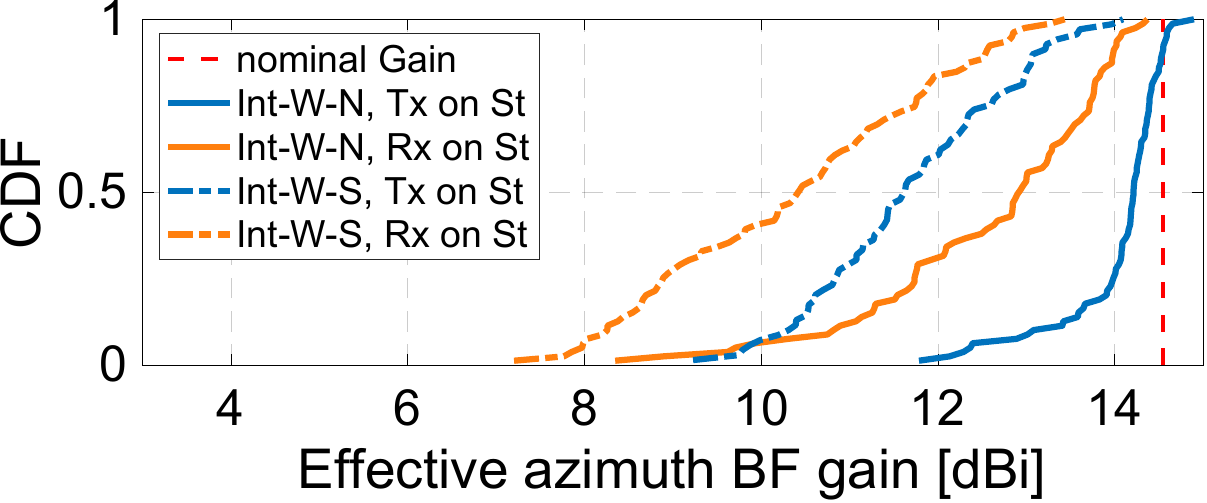}
\vspace{-0\baselineskip}
\label{fig:txrxswap-compare-bfg}}
\hspace{10pt}
\setcounter{subfigure}{3}
\subfloat[]{
\includegraphics[width=0.3\linewidth]{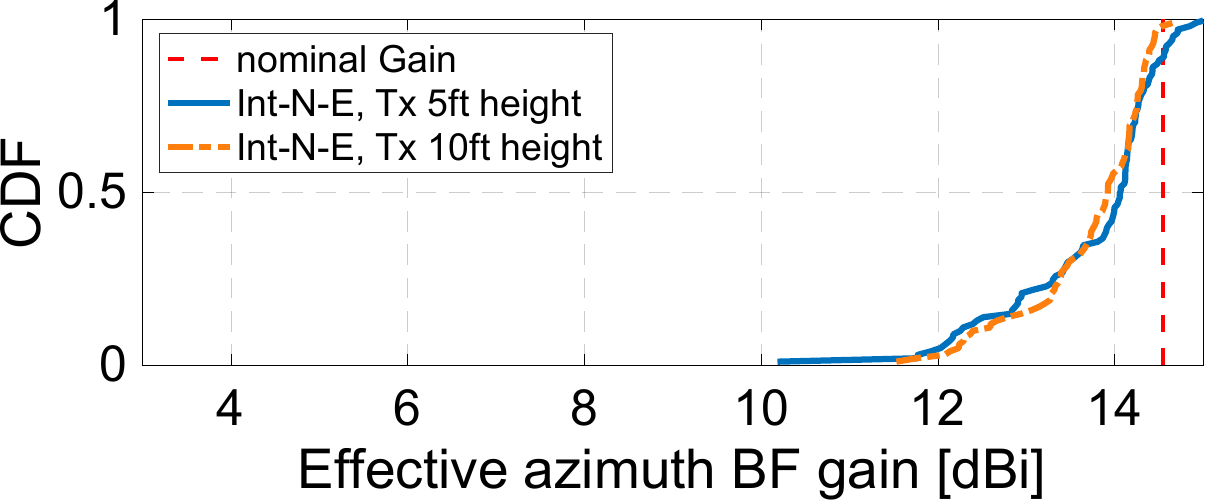}
\vspace{-1\baselineskip}
\label{fig:txheight-compare-bfg}}
\hspace{10pt}
\setcounter{subfigure}{5}
\subfloat[]{
\includegraphics[width=0.3\linewidth]{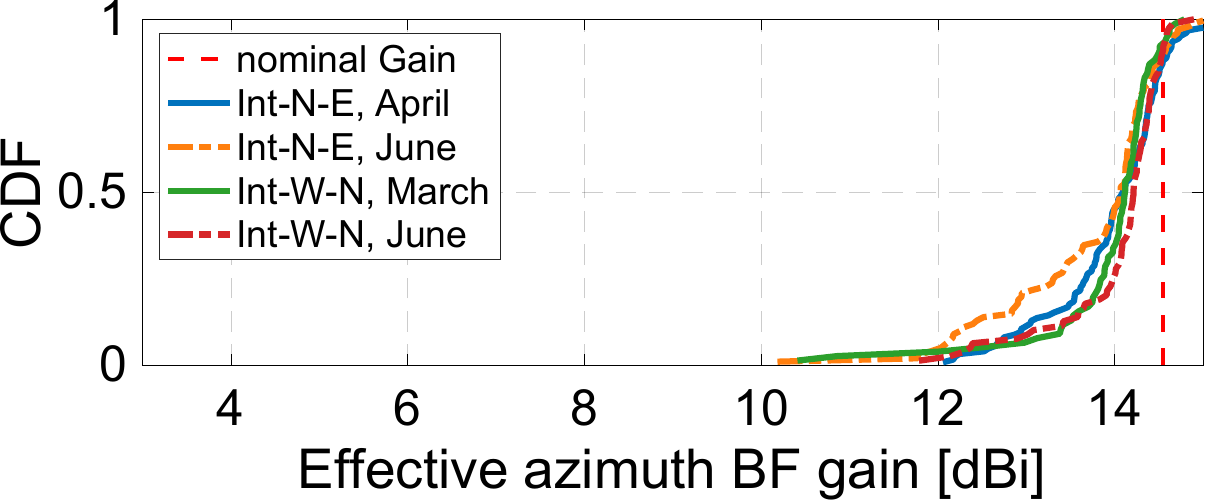}
\vspace{-1\baselineskip}
\label{fig:seasonal-compare-bfg}}
\vspace{\baselineskip}
\setcounter{subfigure}{6}
\subfloat[]{
\includegraphics[width=0.3\linewidth]{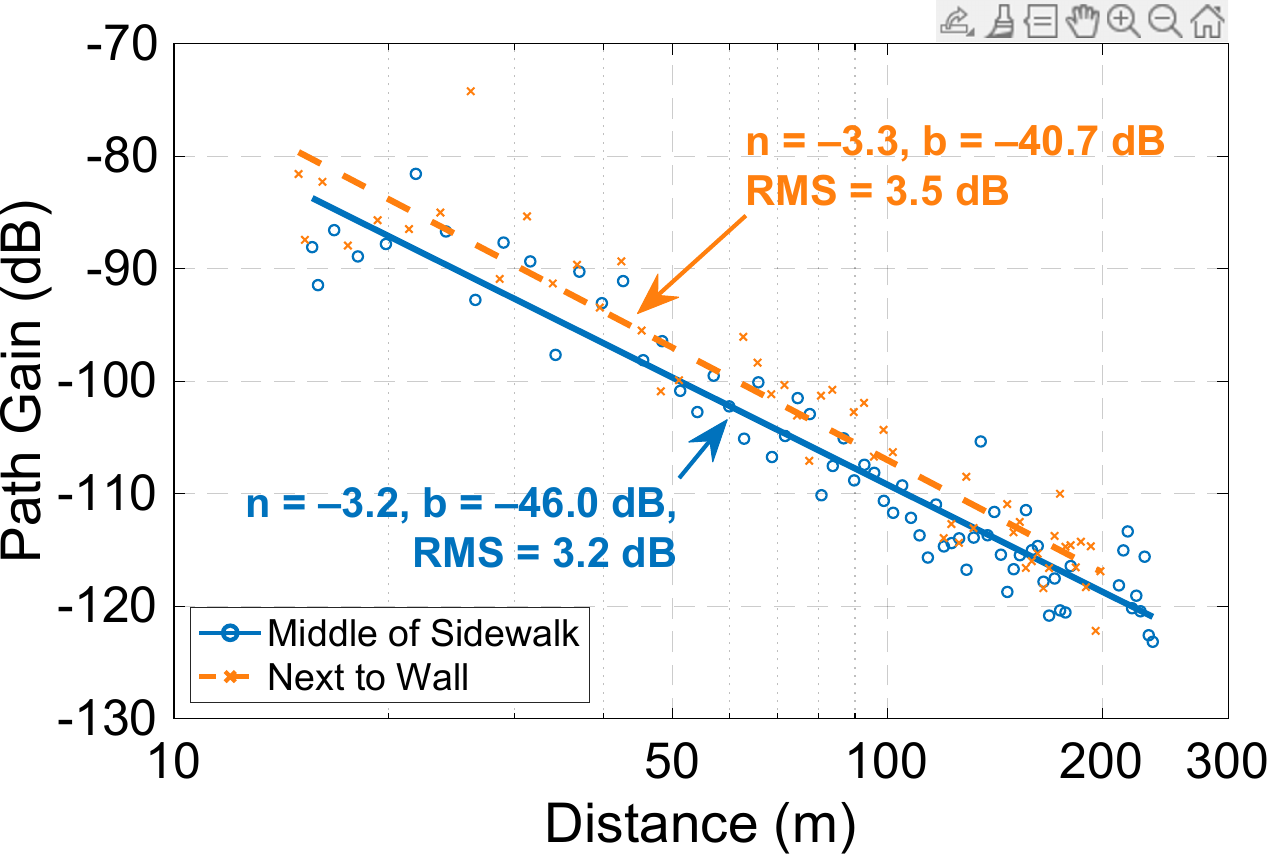}
\label{fig:sidewalkwall-compare-pg}}
\hspace{10pt}
\setcounter{subfigure}{8}
\subfloat[]{
\includegraphics[width=0.3\linewidth]{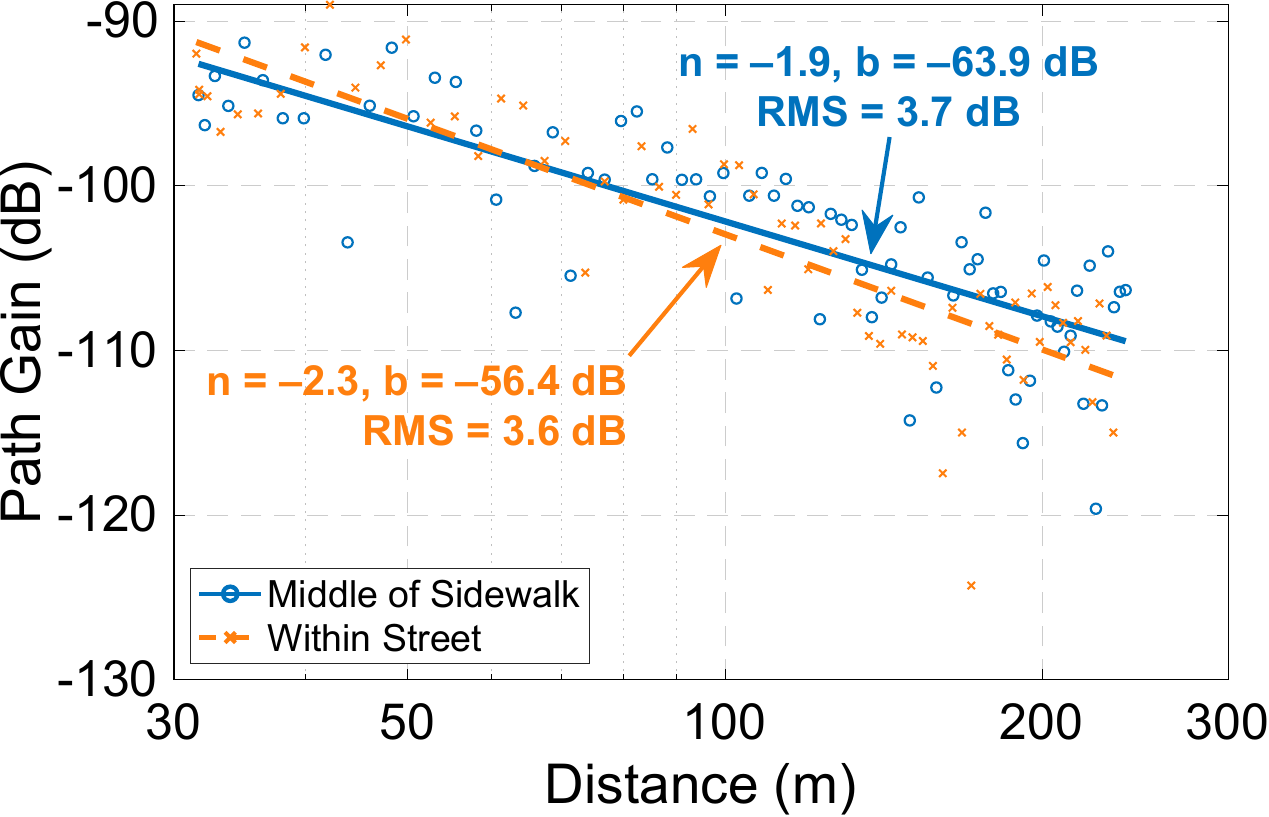}
\label{fig:streetsidewalk-compare-pg}}
\hspace{10pt}
\setcounter{subfigure}{10}
\subfloat[]{
\includegraphics[width=0.3\linewidth]{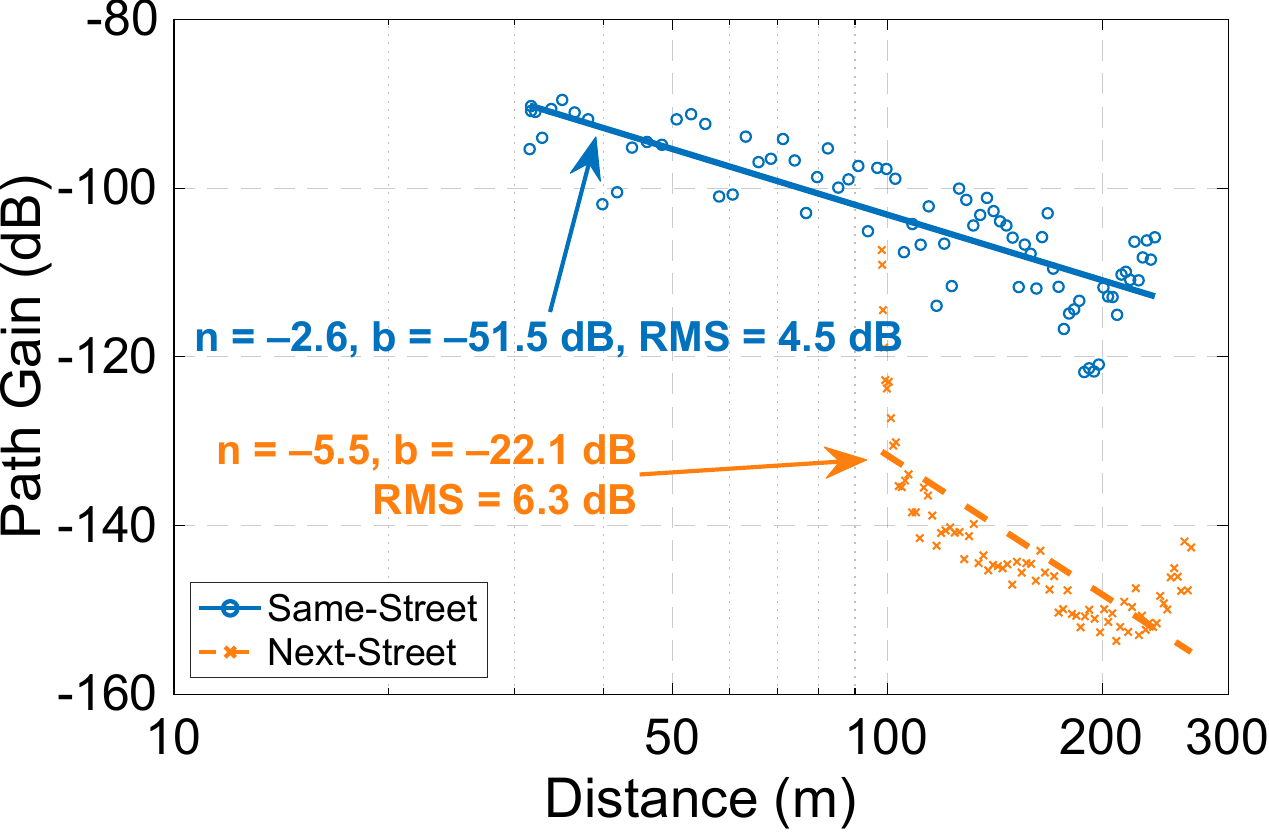}
\label{fig:nextstreet-compare-pg}}
\setcounter{subfigure}{7}
\subfloat[]{
\includegraphics[width=0.3\linewidth]{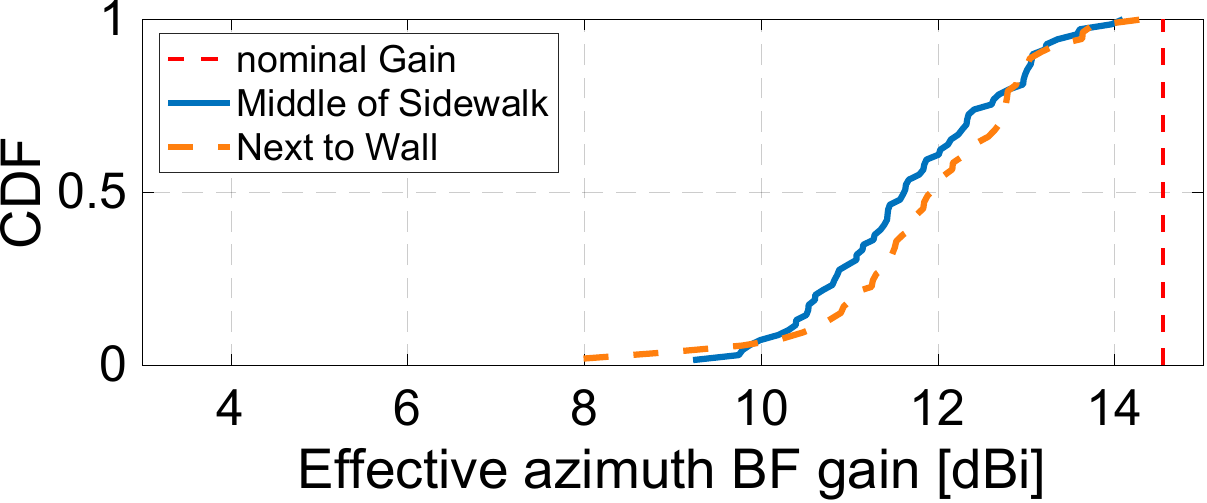}
\label{fig:sidewalkwall-compare-bfg}}
\hspace{10pt}
\setcounter{subfigure}{9}
\subfloat[]{
\includegraphics[width=0.3\linewidth]{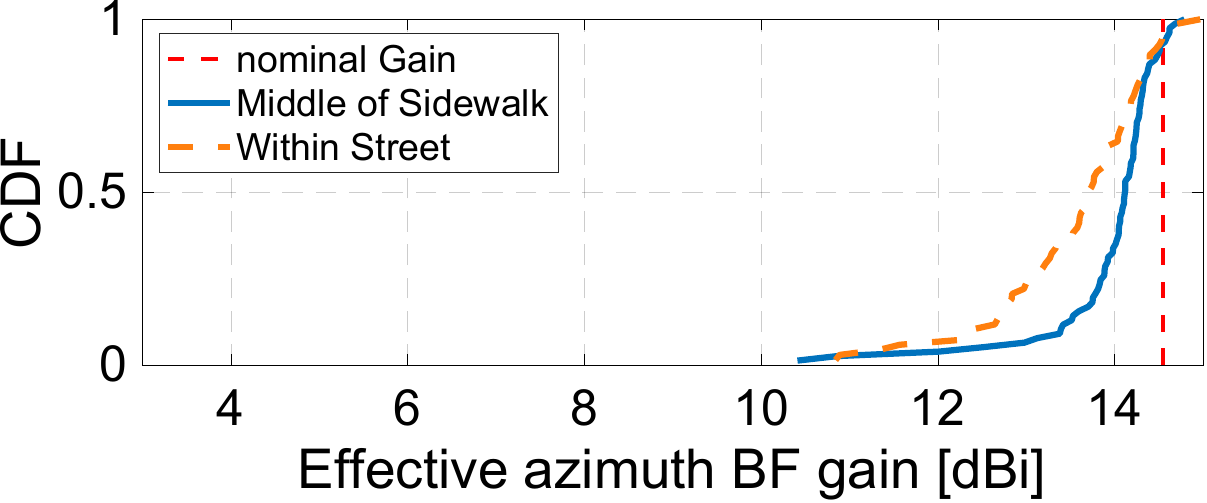}
\label{fig:streetsidewalk-compare-bfg}}
\hspace{10pt}
\setcounter{subfigure}{11}
\subfloat[]{
\includegraphics[width=0.3\linewidth]{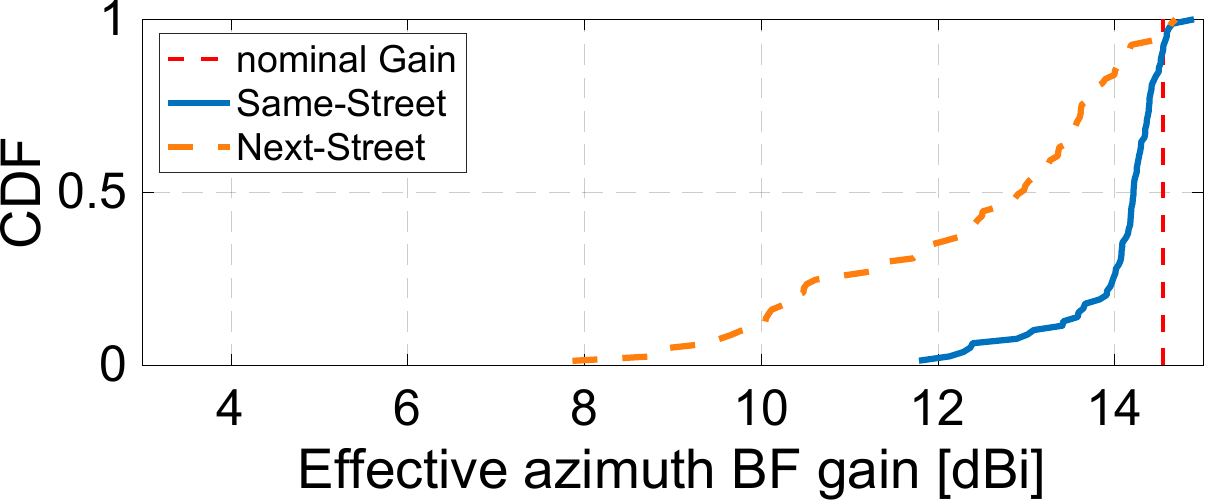}
\label{fig:nextstreet-compare-bfg}}
\vspace{-0.5\baselineskip}
\caption{\addedMK{Path gain and ABG of measurements taken at \textbf{INT} to evaluate potential effects relating to the measurement setup. (a, b) Swapped TX and RX locations; (c, d) TX at different heights; (e, f) measurements taken in different seasons; (g, h) TX located next to the building wall; (i, j) TX located within the street instead of the sidewalk; (k, l) TX located on a sidewalk not adjacent to the RX.}}
\vspace{-0.5\baselineskip}
\label{fig:placement}
\end{figure*}

\subsection{Path Gain and Azimuth Gain}
\label{ssec:meas-metrics}
As described above, the rotating RX with a horn antenna can record the received signal power as a function of both time and the azimuth angle. We now briefly describe the calculation for the \addedMK{three considered performance metrics:}
\begin{enumerate}
    \item \addedMK{The path gain (PG), $G_\text{path}(d)$, which characterizes the signal loss as a function of the Euclidean TX-RX link distance $d$.}
    \item \addedMK{The effective azimuth beamforming gain (ABG), $G_\text{az}(d)$, which quantifies the scattering of the signal brought about by the environment.}
    \item \addedMK{The temporal K-factor, $K(d)$, which describes the relative strength of the time-varying component of the received signal. A larger $K(d)$ signifies are more stable received power over the duration of the experiment.}
\end{enumerate}

The method for calculating these values is described in detail in~\cite{kohli2024outdoor, du2018suburban}. In summary, the instantaneous received power measurement at link distance $d$ and azimuth direction $\phi$, $P(d, \phi)$ may be used to compute $G_\text{path}(d)$ as
\begin{equation}
    G_\text{path}(d) = \frac{1}{2\pi}\left(\int_{\phi=0}^{2\pi} P(d, \phi)d\phi\right)\cdot\frac{1}{P_\text{TX}\cdot G_\text{el}(\theta)}
\label{eq:pg}
\end{equation}

\noindent Where $P_\text{TX}$ is the transmit power of the omnidirectional TX (22\thinspace{dBm}), and $G_\text{el}(\theta)$ is a correction factor to account for the zenith angle $\theta$ due to an elevation mismatch between the TX and RX.

$G_\text{az}(d)$ may be calculated as the maximum received power in azimuth divided by the average received power (i.e., an effective antenna gain in azimuth):
\begin{equation}
    G_\text{az}(d) = \max\nolimits_{\phi}\{P(d, \phi)\}\cdot2\pi\left(\int_{\phi=0}^{2\pi} P(d, \phi)d\phi\right)^{-1}
\label{eq:abg}
\end{equation}

The temporal K-factor is calculated with the method of moments as described in~\cite{greenstein1999moment}.


\section{Validation of the Measurement Method}
\label{sec:validation}
Conducting a \addedMK{well-controlled set of channel measurements within} an urban street canyon environment is challenging, due to \addedMK{potential variation} in the measurement setup and environments. In order to better understand the \addedMK{possible} effects of these factors on the results, we \addedMK{conducted a set of six basic sets of measurements to study the following effects.}


\subsection{Swapped Tx-Rx}
\addedMK{As mentioned in Section~\ref{ssec:meas-environment},} we use the Rx to emulate the BS due to its capability to capture received power across the full 360$^{\circ}$ azimuth plane. Ideally, swapping the placement of the omni-directional Tx antenna and a rotating directional Rx antenna \addedMK{will result} in the same average power as long as the \addedMK{relative elevation gain $G_\text{el}(\theta)$ is compensated for}. 

\addedMK{We collected measurements on \textbf{INT}-W-N and \textbf{INT}-W-S with the Tx and Rx swapped with each other (i.e., the Tx stationary on the \textbf{INT} balcony and the Rx moving along the sidewalk) to verify that the path loss measurements are consistent with the ``standard'' \textbf{INT}-W-N and \textbf{INT-W-S} scenarios.}

Fig.~\ref{fig:placement}\subref{fig:txrxswap-compare-pg} shows the measured path gain values and their fitted lines for the four scenarios to study this effect. The link measurements on the same street but with swapped Tx and Rx locations are similar, where the differences in $n$ (slope) and $b$ (intercept) of the fitted lines are 0.1/0.2 and 2.2/6.7\thinspace{dB} for \textbf{INT}-W-N/\textbf{INT}-W-S, respectively. \addedMK{The difference for \textbf{INT}-W-S is more significant, likely due to the more complex NLOS environmental effects, and we note the fitted lines are still within the RMS fit error at longer Tx-Rx distances.} These results show that swapping the Tx and Rx locations has only minimal effects on the path gain. \addedMK{We therefore use the rotating Rx with better angular resolution to emulate the static BS at the locations in Section~\ref{ssec:meas-environment}.}

\addedMK{Fig.~\ref{fig:placement}\subref{fig:txrxswap-compare-bfg} shows a significant difference in the measured ABG with swapped Rx and Tx, with a median decrease of around 1\thinspace{dB}. As the receiver was placed within the street canyon, it is significantly more likely to receive reflections off closeby structures or the opposite side of the street compared to when the Rx is placed on the relatively open balcony at \textbf{INT}. These effects are reflected by a lower ABG.}

\subsection{Tx Height}
\addedMK{We evaluate the effects of the Tx height by taking two measurements on \textbf{INT}-N-E with the Tx at 5\thinspace{ft} or 10\thinspace{ft} height. Fig.~\ref{fig:placement}\subref{fig:txheight-compare-pg} shows the measured path gain, which differs for the two Tx heights by only {2.5}\thinspace{dB} and {0.8}\thinspace{dB} at the near ($d = 50$\thinspace{m}) and far ($d = 500$\thinspace{m}) ends, respectively. Furthermore, the ABG in Fig.~\ref{fig:placement}\subref{fig:txheight-compare-bfg} shows almost no difference for the two Tx heights. \textbf{INT}-N-E is a busy street with considerable sidewalk clutter, including trees, bus stops, information kiosks, and commercial storefronts. These results demonstrate no significant difference in the measured path loss brought by changes in Tx height even for a street where additional height most likely would have aided the wireless channel by allowing the UE to clear surrounding clutter. Therefore, we use 5\thinspace{ft} as the height for all measurement scenarios.}
\subsection{Seasonal Effects}
\addedMK{The streets surrounding the four measurement locations contain trees and other foliage, typical for NYC and other urban areas. As the measurements were carried out during the period of March--September 2019 and October--December 2020, we conducted measurements on \textbf{INT}-N-E and \textbf{INT}-W-N in two different seasons to observe any potential differences due to the presence of leaves on street foliage. Measurements for the \textbf{INT}-N-E and \textbf{INT}-W-N in Table~\ref{tab:meas-summary} scenarios were collected in summer months, with foliage in full bloom, while \textbf{INT}-N-E-NLe and \textbf{INT}-W-N-NLe in Table~\ref{tab:meas-summary-addl} were collected in the winter months, with trees barren of leaves.}

\addedMK{Fig.~\ref{fig:placement}\subref{fig:seasonal-compare-pg} shows a moderate difference in the measured path gain for the two measurements on \textbf{INT}-W-N, with longer distance links experiencing around 3\thinspace{dB} additional path loss in June compared to March. There is a larger difference observed on \textbf{INT}-N-E, with a reduction in path gain of around 10\thinspace{dB} at link distances beyond 400\thinspace{m}. However, Fig.~\ref{fig:placement}\subref{fig:seasonal-compare-bfg} shows minimal difference in the measured ABG between the different seasons, except for a roughly 1\thinspace{dB} reduction for the \textbf{INT}-N-E scenario with leaves at the 30\textsuperscript{th} percentile. Not every measurement in Table~\ref{tab:meas-summary} was collected in the same season, therefore the aggregate results presented  in Section~\ref{sec:results} include variation brought about by different seasons. The difference in measured path gain between a leaves and no-leaves scenario is similar to that measured in the context of outdoor-to-indoor scenarios with the same measurement equipment~\cite{kohli2024outdoor}.}


\subsection{Tx Proximity to Buildings}
\addedMK{A key component of the measurement setup was to keep the Tx in the middle of the sidewalk as much as possible, mimicking the typical place a person would walk. While generally uniform throughout all the measurement locations, some sidewalks are wider than others, and some had irregular shape along its length. In order to investigate whether the proximity of the Tx to the building adjacent to a sidewalk has any meaningful effect on the measured path gain, we conducted the measurement \textbf{INT}-W-S-Wall in Table~\ref{tab:meas-summary-addl} where we placed the Tx as close to the building wall as possible (on the order of centimeters). Fig.~\ref{fig:placement}\subref{fig:sidewalkwall-compare-pg} shows a roughly 3\thinspace{dB} increase in path gain when the Tx is placed next to the wall, which is within the error of the two fitted models. Therefore, we find no significant difference in the measured path gain as a result of changes in the lateral position of the Tx on the sidewalk; for the measurements in Table~\ref{tab:meas-summary}, the Tx was never placed as close to the building wall in any case.}

\subsection{Tx on Sidewalk or on Street}
\addedMK{The \textbf{INT}-W-N sidewalk is located behind a row of reverse-in parking spaces, which are typically fully occupied during a workday, and other sidewalks are similarly located behind rows of parked cars. To further validate the choice of Tx location in the middle of the sidewalk, we consider the scenario \textbf{INT}-W-N-St where the Tx is placed within the street itself. Fig.~\ref{fig:placement}\subref{fig:streetsidewalk-compare-pg} shows a minimal difference between the two conditions, which is within the model fit error for the entire distance measured. There is a minor difference in ABG shown by Fig.~\ref{fig:placement}\subref{fig:streetsidewalk-compare-pg}, roughly 1\thinspace{dB} at the 30\textsuperscript{th} percentile. These results demonstrate a similar effect as did the scenarios with differing Tx heights - the sidewalk clutter, including parked cars and sidewalk fixtures, do not seem to have a significant effect on the measured path gain.}

\subsection{Use of Adjacent Sidewalks}
\addedMK{Lastly, we evaluated the effect of measuring on a sidewalk that is not adjacent to the Rx. We consider the \textbf{INT}-W-N2 scenario in Table~\ref{tab:meas-summary-addl} which was collected one block north of the \textbf{INT} location, as indicated in Fig.~\ref{fig:locmaps}\subref{fig:loc-int}. While this scenario is VNLOS as with others in the measurement set, \textbf{INT}-W-N2 experiences particularly heavy blockage from the city block separating the Tx and Rx. As expected, Fig.~\ref{fig:placement}\subref{fig:nextstreet-compare-pg} shows a severe path gain decrease over the \textbf{INT}-W-N scenario, in excess of 40\thinspace{dB} at 200\thinspace{m} link distance. The ABG is degraded at the median by 1\thinspace{dB} compared to the same-street sidewalk, and even further at the lower percentiles.}

\addedMK{The path gain measurements demonstrate that for the nearest locations to the Rx in the \textbf{INT}-W-N2 scenario, namely at the street corner closest to VLOS with the Rx, the path loss compares to the same street scenario. The path loss then decreases rapidly, before entering a slower decrease regime. Furthermore, at longer distances, as the Tx reaches the far end of the street block, the path gain begins to increase. The propagation mechanism in this scenario is decidedly complex and best modeled via diffuse scatter~\cite{chizhik2023accurate} rather than an urban canyon model, and as a result, we do not consider further measurements of this type in this dataset. A closer investigation of mmWave signal propagation around street corners is a subject of our current and future research.}

\addedMK{Overall, we believe that these measurements collectively justify and validate the measurement methodology set out in Section~\ref{sec:meas} and used in the remainder of the paper. Independent of the goal of validating the methodology, the effects studied in this section also provide insight on some potential impacts (or lack thereof) on the mmWave channel brought about by changes in UE location or the street sidewalk environment.}

\section{Measurement Results and Analysis}
\label{sec:results}
\addedMK{In this section, we present and discuss a series of measurement results collected from the measurement scenarios in Table~\ref{tab:meas-summary}. In particular, we compute path gain models under certain conditions, such as the location or whether or not a measured link was in VLOS or VNLOS. We also look in detail at how the angle of arrival (AoA) recorded at the rotating Rx changes as the Tx moves down a street sidewalk.}

\begin{figure*}[t]
\centering
\vspace{-1\baselineskip}
\subfloat[]{
\includegraphics[width=0.45\linewidth]{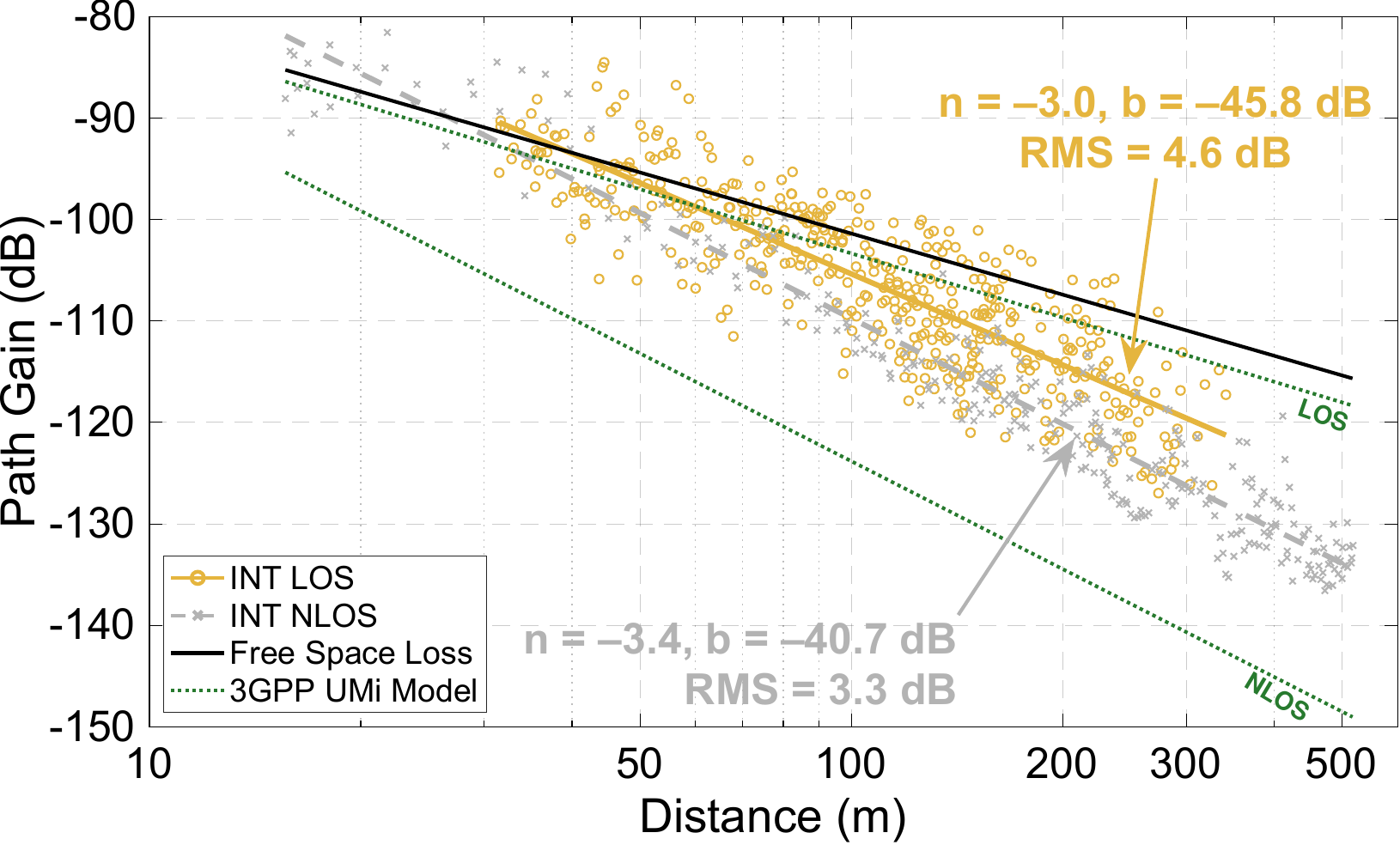}
\vspace{-1\baselineskip}
\label{fig:pg-int}}
\hspace{10pt}
\subfloat[]{
\includegraphics[width=0.45\linewidth]{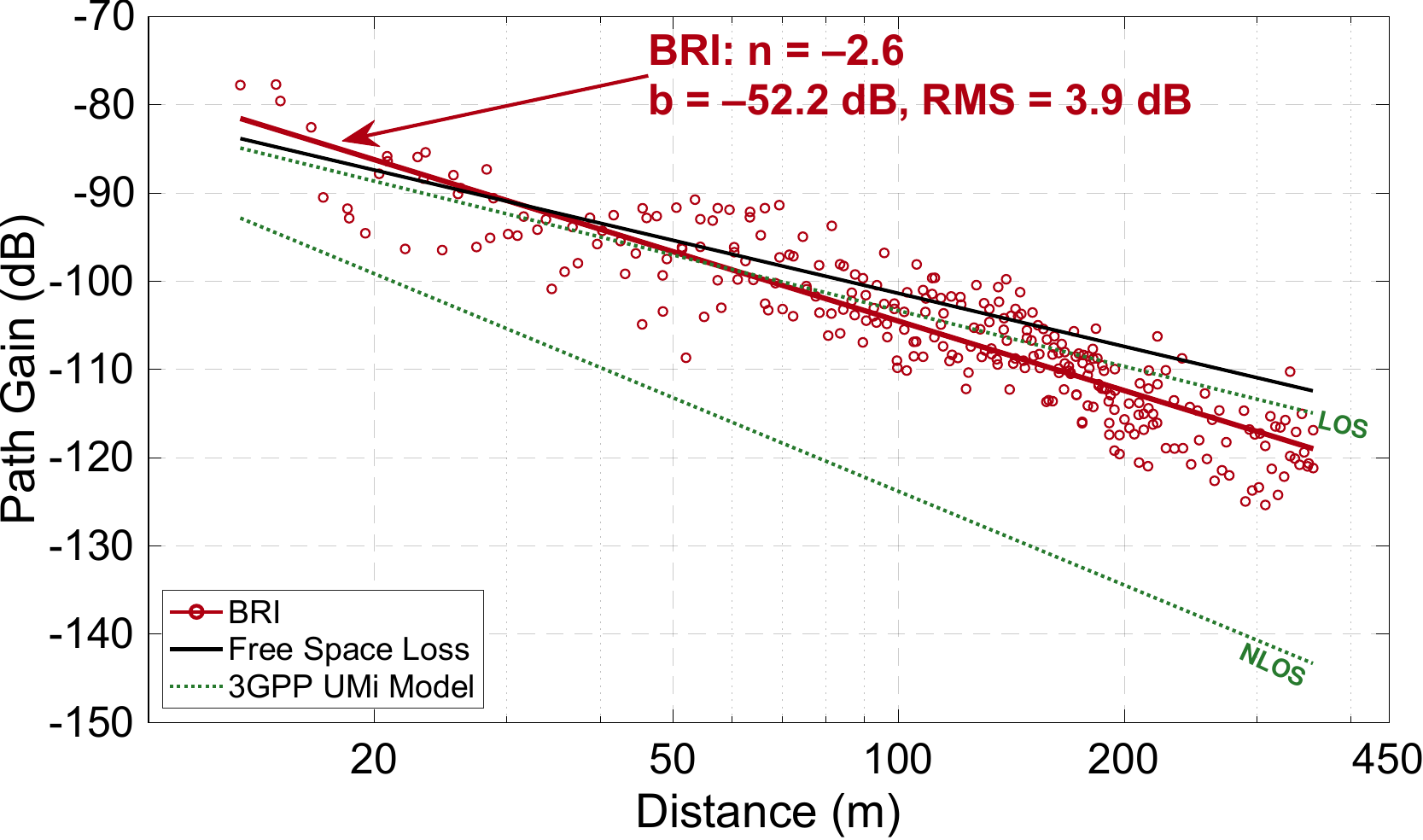}
\vspace{-1\baselineskip}
\label{fig:pg-bri}}
\\
\subfloat[]{
\includegraphics[width=0.45\linewidth]{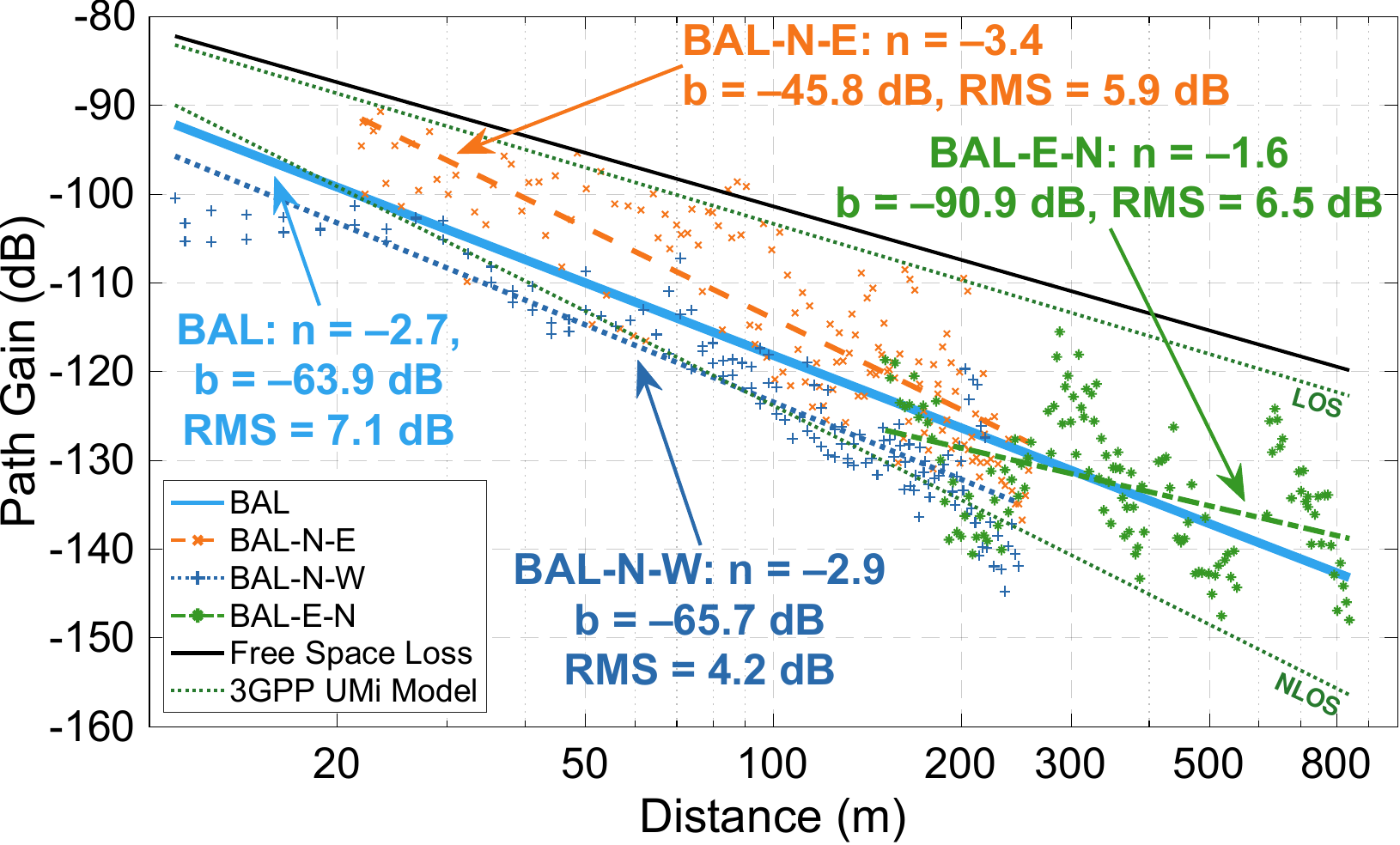}
\vspace{-1\baselineskip}
\label{fig:pg-bal}}
\hspace{10pt}
\subfloat[]{
\includegraphics[width=0.45\linewidth]{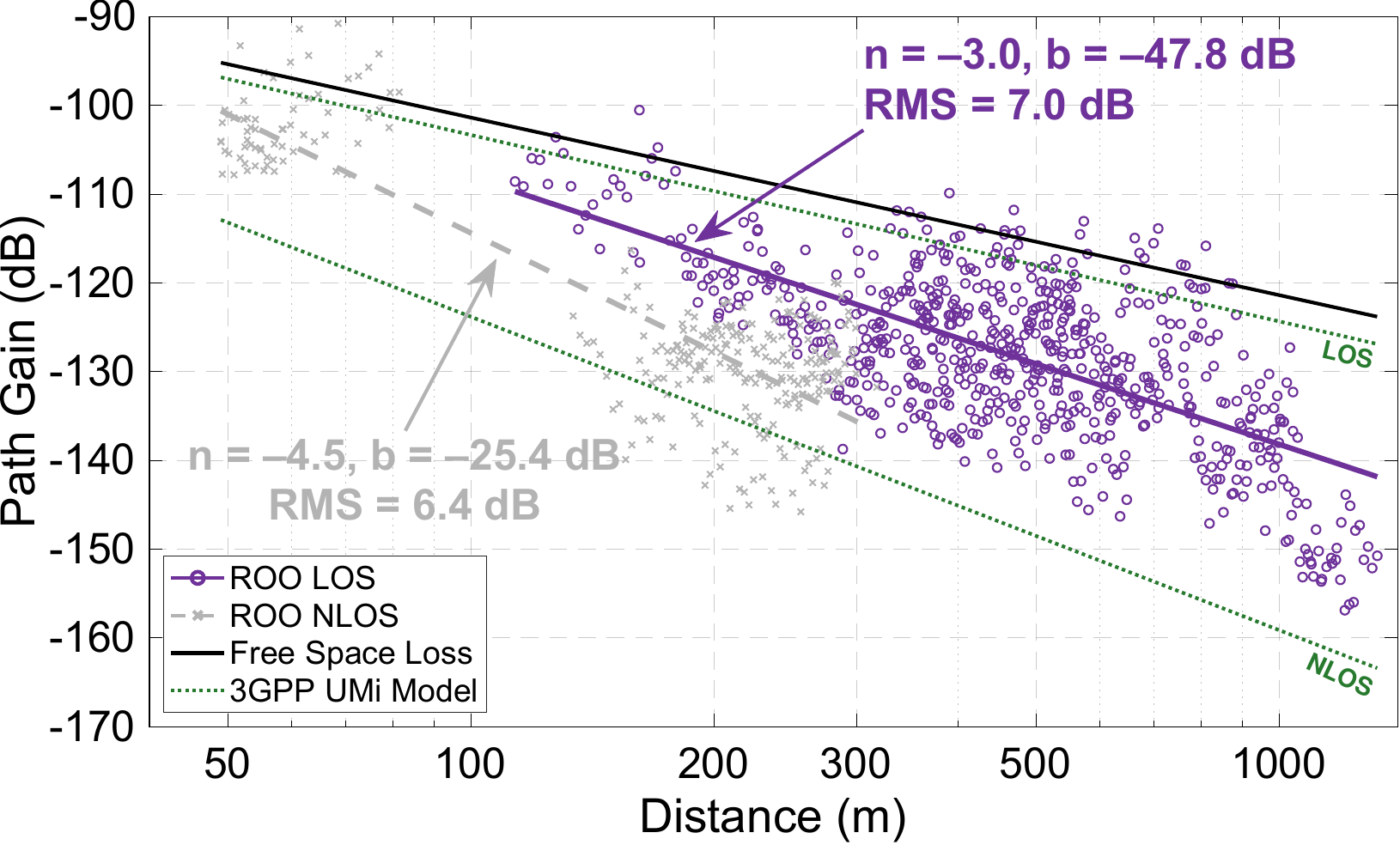}
\vspace{-1\baselineskip}
\label{fig:pg-roo}}
\vspace{-0.5\baselineskip}
\caption{\addedMK{Path gain models for measurements taken at the four Rx locations. (a) \textbf{INT} VLOS and NVLOS; (b) \textbf{BRI}; (c) \textbf{BAL} and the three scenarios provided separately; (d) \textbf{ROO} LOS and NLOS.}}
\vspace{-1\baselineskip}
\label{fig:pg}
\end{figure*}

\subsection{Path Gain}
\label{ssec:results-pg}
\addedMK{We compute path gain models for the four Rx locations, \textbf{INT}, \textbf{BRI}, \textbf{BAL}, and \textbf{ROO}. Due to the large number of VNLOS measurements conducted at \textbf{INT} and \textbf{ROO}, we split the measurements taken at those locations into two sets for VLOS and VNLOS for each location. The VLOS and VNLOS links correspond to the location of the solid and dashed arrows, respectively, shown in Figs.~\ref{fig:locmaps}\subref{fig:loc-int} and \ref{fig:locmaps}\subref{fig:loc-roo}. Fig.~\ref{fig:pg} presents the path loss models in each case as a function of Euclidean Tx-Rx link distance, along with LOS and NLOS models derived from the 3GPP UMi~\cite{3gpp_los_nlos}.}

\subsubsection{INT-VLOS and INT-VNLOS}
\addedMK{Fig.~\ref{fig:pg}\subref{fig:pg-int} shows the path loss models computed for all VLOS and VNLOS measurements collected at \textbf{INT}. These models show that the NLOS condition experiences a faster decrease in path gain as a function of distance compared to LOS, with $n = -3.0$ for VLOS and $n = -3.4$ for VNLOS. We note that a significant number of VLOS and some VNLOS measurements at shorter distances are above the free space path loss (FSPL). This effect was observed at all locations, but most frequently at \textbf{INT}, and suggests that a very regular urban street canyon environment can provide higher gain to the mmWave channel than generalized models suggest. Likewise, the VNLOS measurements demonstrate around 15\thinspace{dB} higher dB path gain than the NLOS 3GPP UMi model for the entire range of Tx-Rx link distance.}

\subsubsection{BRI}
\addedMK{The path gain model for the measurements taken at \textbf{BRI} is provided in Fig.~\ref{fig:pg}\subref{fig:pg-bri}. This measurements at \textbf{BRI} demonstrated a higher path gain than those at other locations; this is possibly due to the highly regular street canyon and the location of the Rx in the middle of the street, providing a highly symmetric environment. Like at ~\textbf{INT}, some measurements at \textbf{BRI} experience path gain in excess of not only the 3GPP UMi LOS model but also free space, but the overall path gain model lies below the UMi LOS model for the majority of measured distances.}

\subsubsection{BAL}
\addedMK{The path gain models for the \textbf{BAL} location and the three scenarios in Table~\ref{tab:meas-summary} and Fig.~\ref{fig:locmaps}\subref{fig:loc-bal} are shown in Fig.~\ref{fig:pg}\subref{fig:pg-bal}. Each scenario is provided separately as they each have a unique feature that likely contributes significantly to the measured path gain.}

\addedMK{\textbf{BAL}-N-E and \textbf{BAL}-N-W are two sides of a street located next to an open urban park. Therefore, the urban canyon effect is lost as one side of the street does not contain a row of buildings. \textbf{BAL}-N-W is significantly impacted by environment, with its path gain model lying very close to the 3GPP UMi NLOS model. The difference between \textbf{BAL}-N-E and \textbf{BAL}-N-W averages around 10\thinspace{dB}, close to the difference between opposite sidewalks \textbf{INT}-W-S and \textbf{INT}-W-N in Fig.~\ref{fig:placement}\subref{fig:txrxswap-compare-pg}, but the models at \textbf{BAL} show around 10\thinspace{dB} lower path gain than \textbf{INT} over equivalent Tx-Rx link distances. This same trend is seen in the overall model for \textbf{BAL} compared to \textbf{INT}-VLOS, suggesting that the loss of the urban canyon effect corresponds to around 10\thinspace{dB} lower path gain.}

\subsubsection{ROO-VLOS and ROO-VNLOS}
\addedMK{Fig.~\ref{fig:pg}\subref{fig:pg-roo} shows the path loss models for the VLOS and VNLOS measurements from \textbf{ROO}. Compared to \textbf{INT}, there is a more pronounced difference between LOS and NLOS at 300\thinspace{m} Tx-Rx distance, with around 5\thinspace{dB} greater difference in path gain at \textbf{ROO}. The NLOS measurements at \textbf{ROO} are closer to the 3GPP UMi NLOS model, with some measurement locations showing in excess of 10\thinspace{dB} lower path gain than this model would predict.}

\begin{figure}[t]
\centering
\includegraphics[width=0.9\columnwidth]{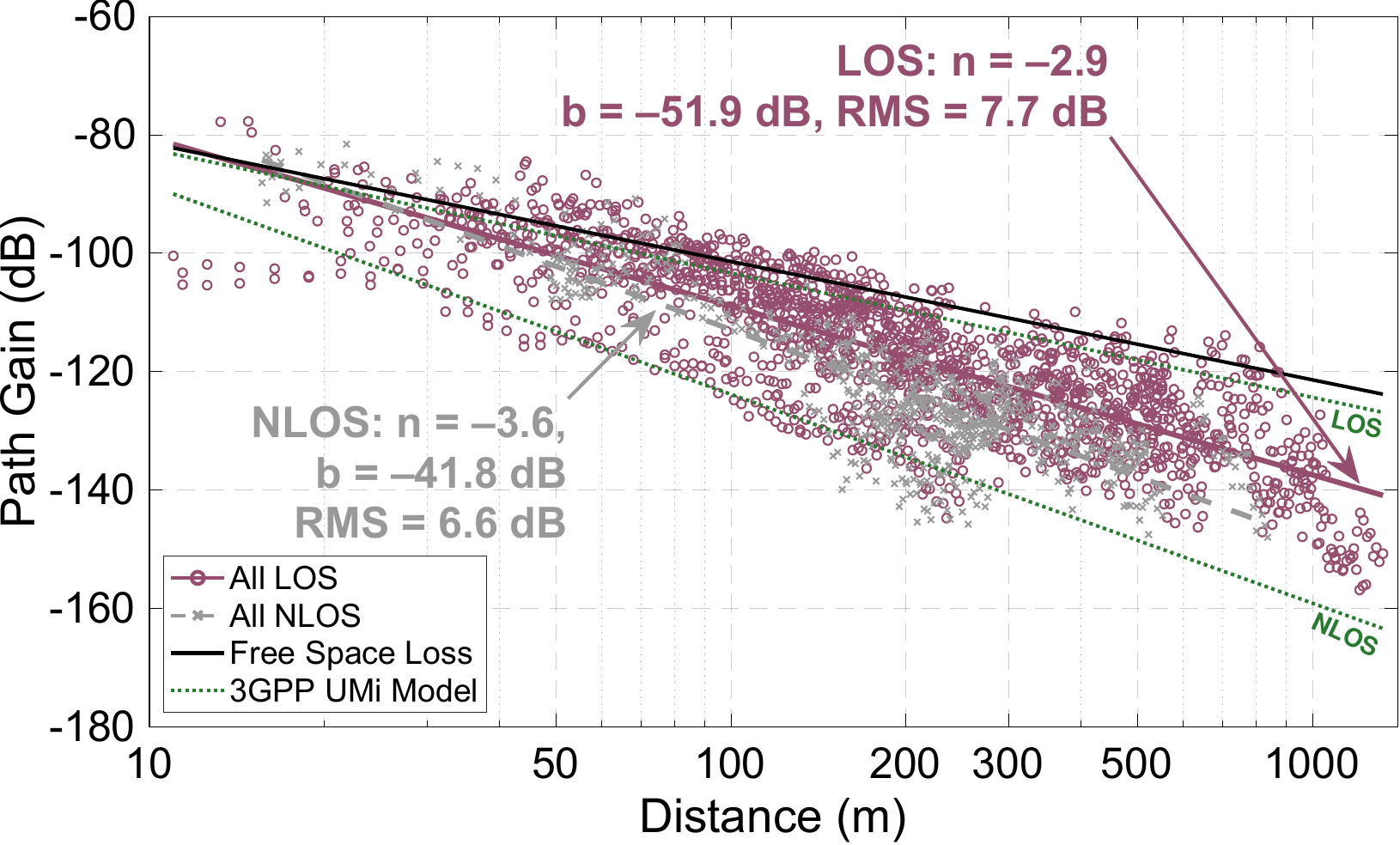}
\caption{\addedMK{Path gain models for all LOS and NLOS measurements at the four Rx locations combined.}}
\label{fig:all-pg}
\vspace{-1\baselineskip}
\end{figure}

\addedMK{Both the VLOS and VNLOS path gain models show a significantly larger RMS error than their counterparts at \textbf{INT}. This can be explained by the measurement environment at \textbf{ROO}, which is significantly more complex than \textbf{INT}, including the presence of an elevated segment of the NYC Subway and an irregular street layout in the vicinity, atypical of the NYC street grid. The large variation in measurement environment for the different scenarios at \textbf{ROO} is directly reflected in Fig.~\ref{fig:pg}\subref{fig:pg-roo}, where the VLOS path gain measurements at a given Tx-Rx distance can differ by over 30\thinspace{dB}, with some above FSPL and others close to even the 3GPP UMi NLOS model.}

\begin{figure*}[t]
\centering
\vspace{-1\baselineskip}
\subfloat[]{
\includegraphics[width=0.3\linewidth,trim={0cm 2.5cm 1.5cm 3.2cm},clip]{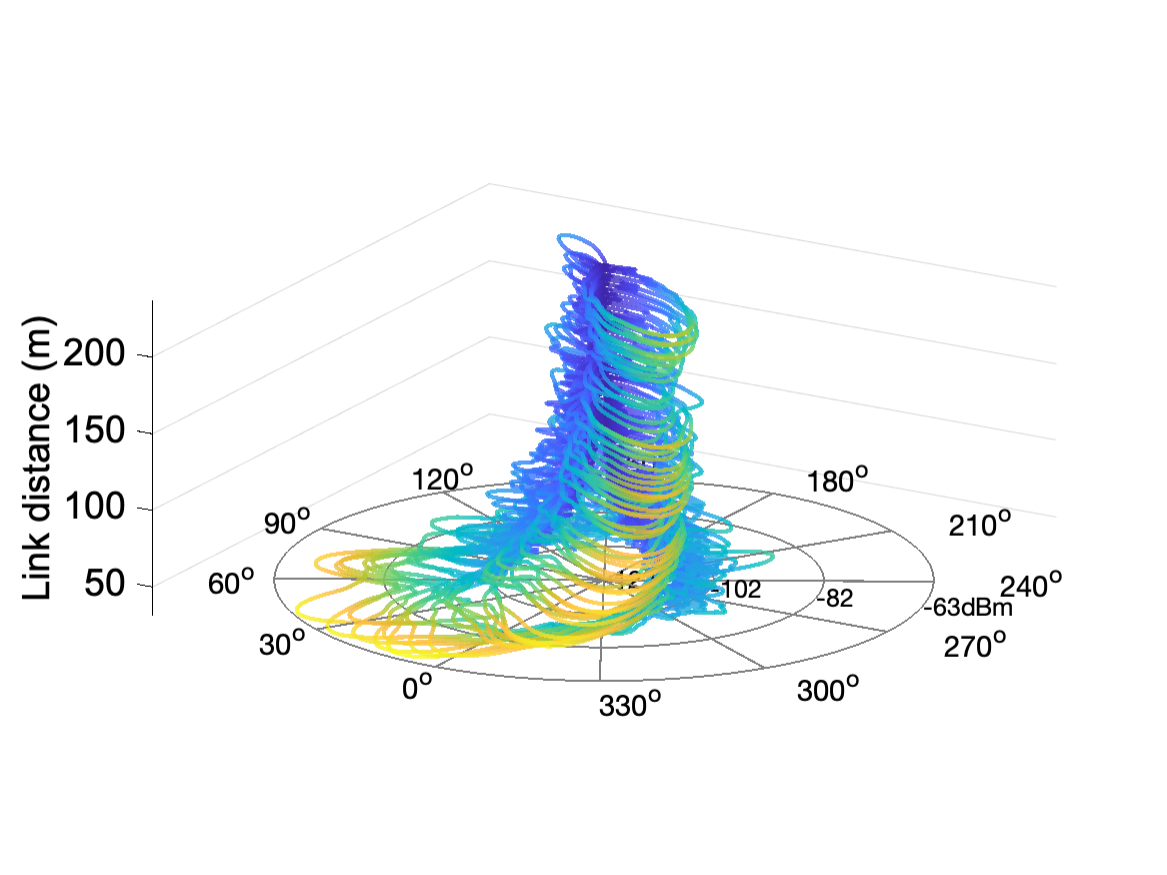}
\label{fig:stack_LOS}}
\hspace{10pt}
\subfloat[]{
\includegraphics[width=0.3\linewidth,trim={0cm 2.5cm 1.5cm 3.2cm},clip]{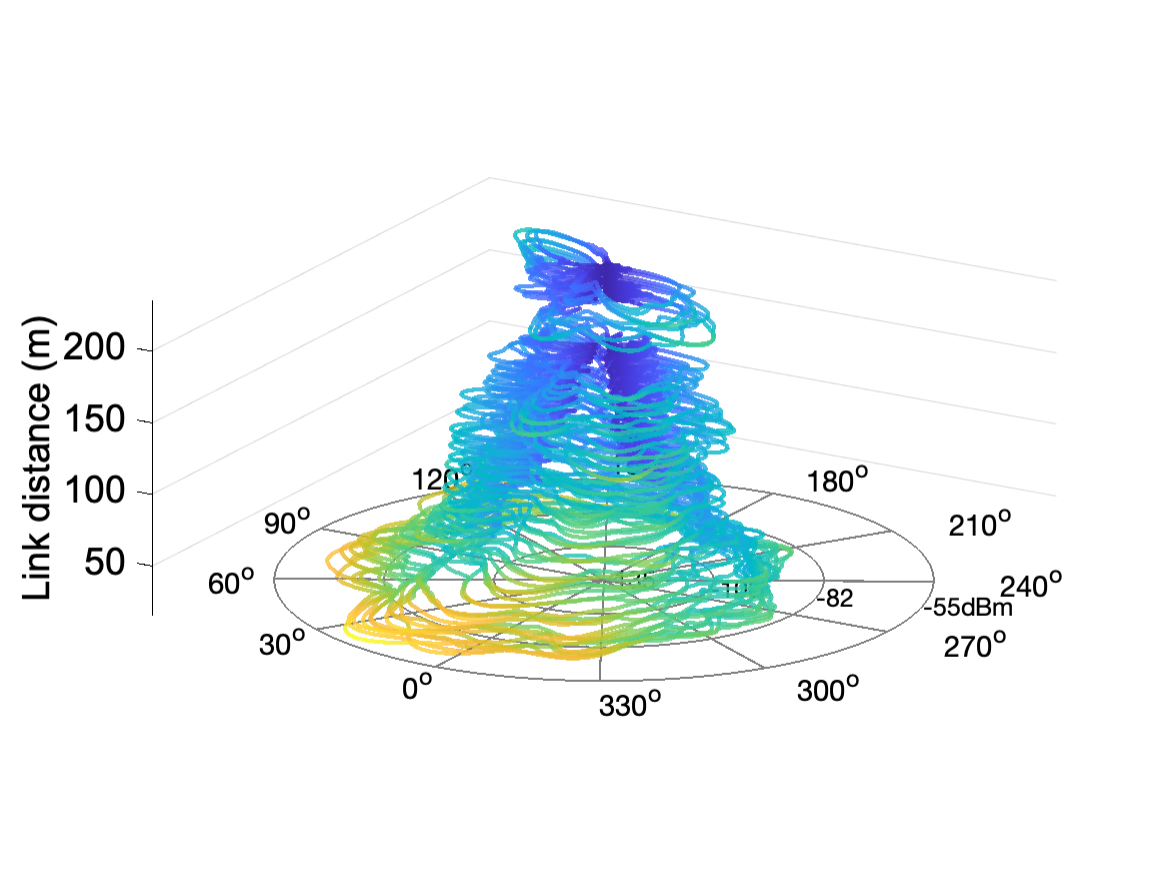}
\label{fig:stack_NLOS}}
\hspace{10pt}
\subfloat[]{
\includegraphics[width=0.3\linewidth,trim={0cm 2.5cm 1.5cm 3.2cm},clip]{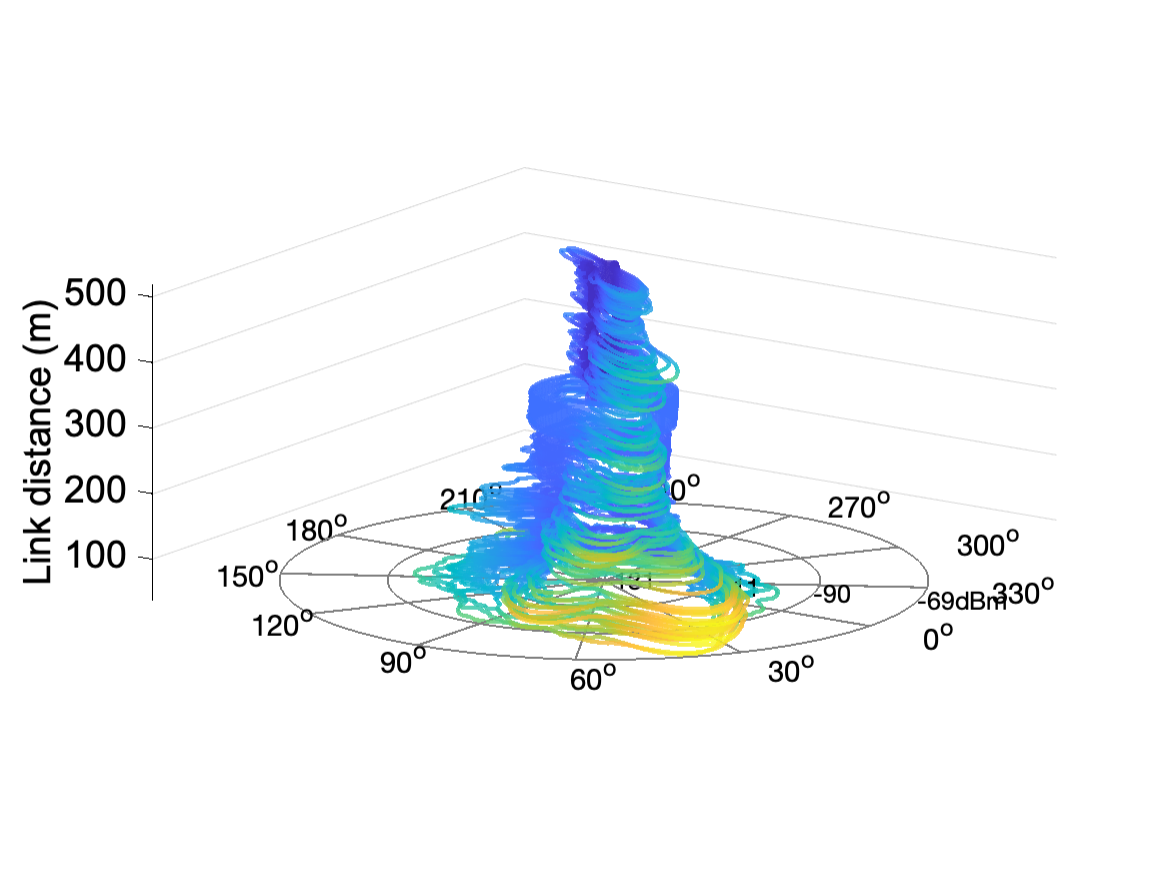}
\label{fig:stack_NLOS2}}
\vspace{-0\baselineskip}
\caption{\addedMK{Power angular spectra against link distance for (a) \textbf{INT}-W-N (LOS); (b) \textbf{INT}-W-S (NLOS); (c) \textbf{INT}-N-W (NLOS)}}
\vspace{-0.5\baselineskip}
\label{fig:stack_spectra}
\end{figure*}

\subsubsection{All Locations}
\label{sssec:results-pg-allloc}
\addedMK{Lastly, we group together all VLOS and VNLOS measurements from the four Rx locations in Fig.~\ref{fig:all-pg}. The path gain models for the LOS and NLOS groups lie  between the 3GPP UMi LOS and NLOS models, with around 10\thinspace{dB} difference between the two at longer distances. This difference is significantly smaller than the  $>30$\thinspace{dB} difference between the 3GPP UMi models.}

\addedMK{TR 38.901, the report which defines the UMi models, does not provide a strict definition of NLOS. Therefore, it is possible that our VLOS/VNLOS conditioning provides an insufficiently strong boundary between these two types of links. However, this does not explain the fact that the individual path gain measurements themselves are located in between the two UMi models, not just the empirical path gain models. Instead, it is more likely that any measurement has a mixed LOS and NLOS character. For example, the NLOS measurements on INT-N-W only start around 150\thinspace{m} link distance because the corner of the building north of the Rx location begins to obscure visual sight of the Rx. However, the blockage caused by this building corner does not preclude the ability for the signal to propagate via reflection from on the other side of the street canyon, or diffraction around the corner~\cite{chizhik2023accurate}. Therefore, it could be said that such locations are of a mixed LOS and NLOS character. We study these alternate propagation mechanisms in more detail in Section~\ref{ssec:results-aoa}.}

\addedMK{This property is considered within TR 38.901, which also provides a LOS probability model as a function of the two-dimensional Tx-Rx link distance $d_\text{2D}$ (i.e., ignoring the vertical offset). This model produces $\rho_\text{LOS} < 0.1$ for $d_\text{2D} > 190\thinspace{m}$. Assuming that the Rx is at the highest measured location (\textbf{ROO}, 48\thinspace{m}), this corresponds to 196\thinspace{m} Tx-Rx link distance. However, it can be seen from Fig~\ref{fig:all-pg} that there are a larger number of measurements closer to the 3GPP UMi LOS model than the NLOS model beyond this distance. To surmise, the UMi path gain models are a good upper and lower bound for our dataset. The LOS UMi model is roughly 10\thinspace{dB} more optimistic than our empirical LOS model, and the NLOS UMi model is roughly 10\thinspace{dB} more pessimistic than our empirical NLOS model. Furthermore, the difference between the empirical LOS and NLOS models is surprisingly low, with only 10\thinspace{dB} difference at longer link distances, and even lower at shorter distances.}

\subsection{Angle of Arrival for NLOS Scenarios}
\label{ssec:results-aoa}




\addedMK{In this subsection, we evaluate the \textbf{INT}-W-S and \textbf{INT}-N-W scenarios to better understand the angle of arrival (AoA) of the signal at the Rx, as this can express propagation mechanisms within the urban canyon.} \addedMK{The rotating Rx measures power $P(d, \phi)$ as a function of Tx-Rx distance $d$ and azimuth angle $\phi$, as described in Section~\ref{ssec:meas-metrics}. By averaging $P(d, \phi)$ over all $\sim$40 turns completed by the Rx during a measurement, we obtain the power angular spectrum (PAS), $\bar{P}(d, \phi)$, describing the power received across the azimuth.}

\begin{figure}[t]
\centering
\vspace{-1\baselineskip}
\subfloat[]{
\includegraphics[width=0.47\columnwidth]{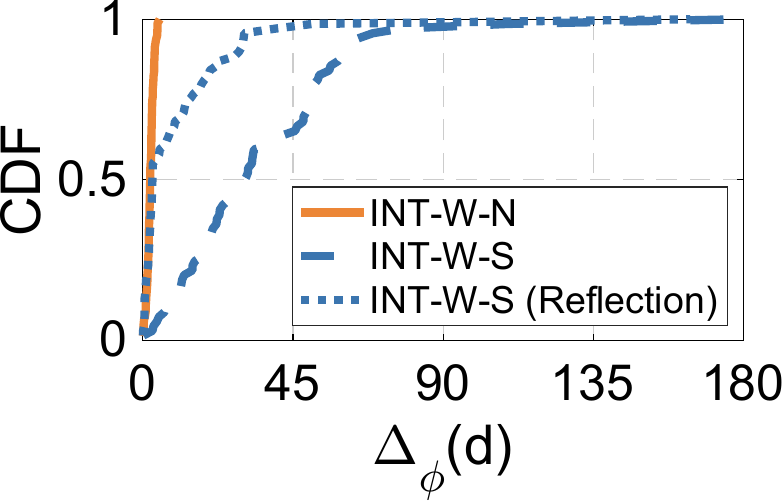}
\label{fig:aoi-intw}}
\subfloat[]{
\includegraphics[width=0.47\columnwidth]{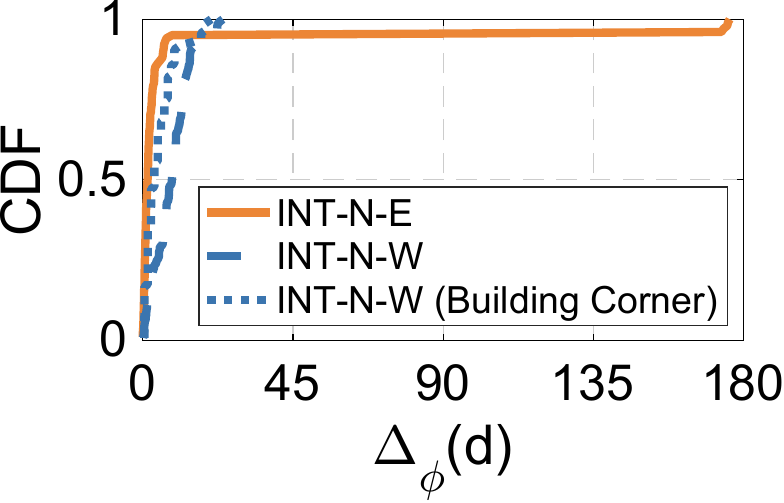}
\label{fig:aoi-intn}}
\\
\vspace{-0.5\baselineskip}
\subfloat[]{
\includegraphics[width=0.59\columnwidth]{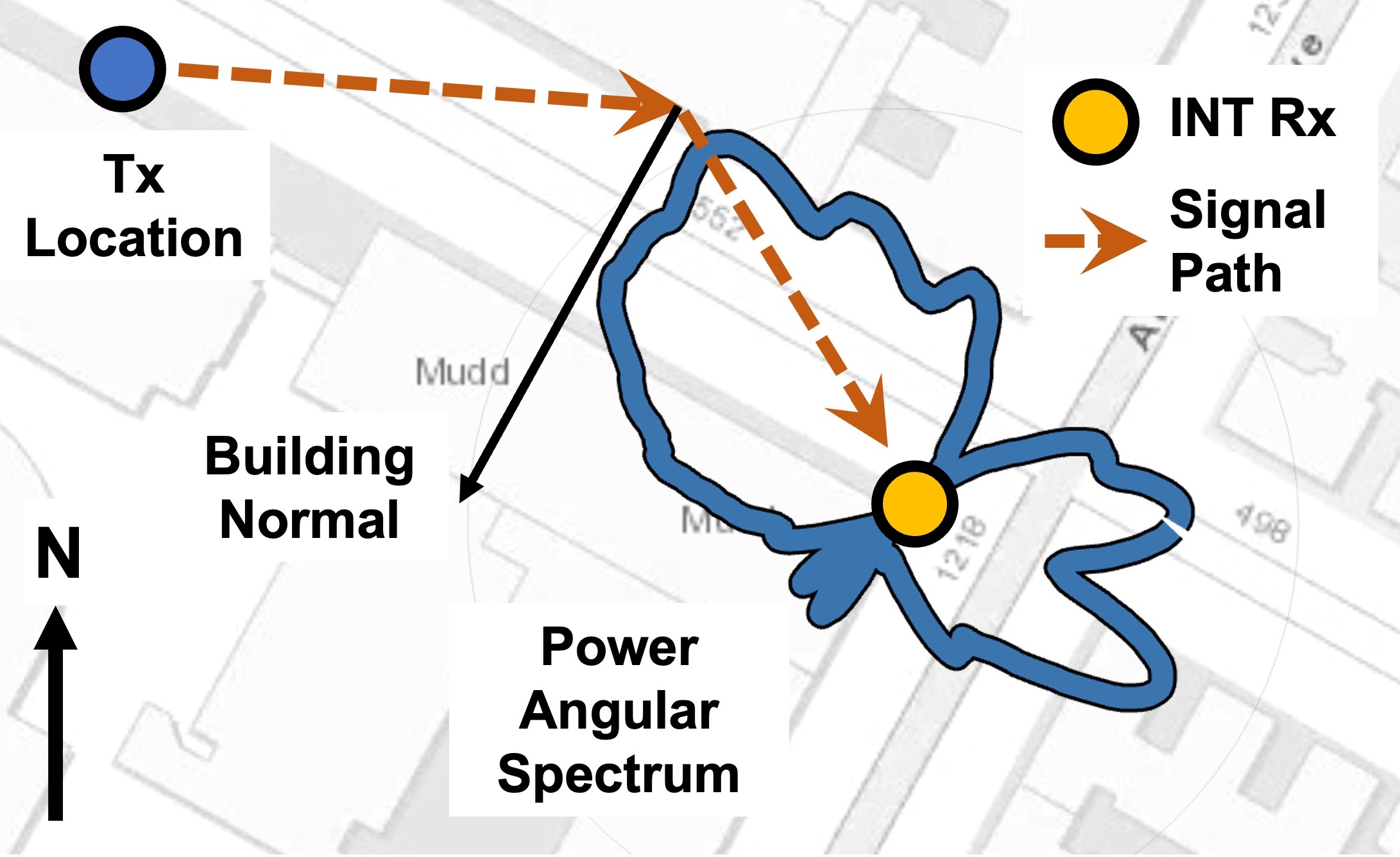}
\label{fig:aoi-intws-pas}}
\subfloat[]{
\includegraphics[width=0.37\columnwidth]{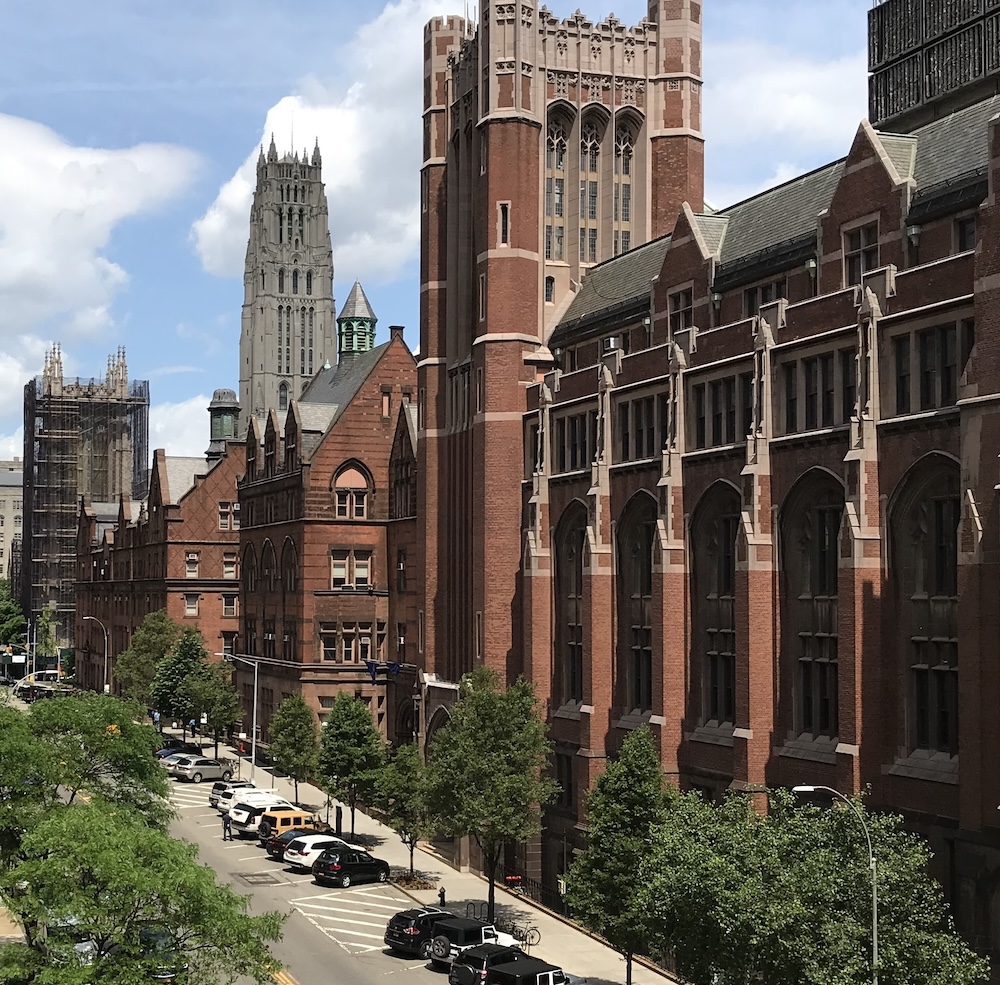}
\label{fig:aoi-intws-opposite}}
\vspace{-0.5\baselineskip}
\caption{\addedMK{AoA-AoI angular deviation $\Delta_\phi(d)$ for (a) \textbf{INT}-W-N and \textbf{INT}-W-S; (b) \textbf{INT}-N-E and \textbf{INT}-N-W. (c) Example reflective signal propagation path for measurement taken on \textbf{INT}-W-S; (d) Building opposite the Rx at \textbf{INT}.}}
\vspace{-1\baselineskip}
\label{fig:aoi}
\end{figure}

\addedMK{By ``stacking'' the recorded PAS at each measured Tx-Rx distance $d$, we can observe visually how the signal propagation changes as the Tx moves away from the Rx. Fig.~\ref{fig:stack_spectra} shows the stacked PAS for \textbf{INT}-W-N, \textbf{INT}-W-S, and \textbf{INT}-N-W. The stacked PAS for the LOS \textbf{INT}-W-N shows the peak power (yellow highlighted region) swing from around 0$^\circ$ to 315$^\circ$. Given 0$^\circ$ represents true north, this matches very closely to street geometry. However, the stacked PAS for the NLOS \textbf{INT}-W-S shows a less clear change in peak received power as the Tx moves away from the Rx. Visually, there is a less well-defined peak, and for larger distances, peaks can be seen pointing in more than one direction. This suggests the presence of a relatively strong environmental reflective path. Indeed, geometrically, these angles correspond to a reflection received off the buildings across the street from the Rx.}

\addedMK{A similar but less drastic effect is seen for the predominantly NLOS \textbf{INT}-N-W sidewalk. Compared to the other side of the street (\textbf{INT}-N-E, not shown in Fig.~\ref{fig:stack_spectra}), the peaks do not follow the direct angle from Tx to Rx.}

\addedMK{These qualitative results can be quantified by considering the deviation $\Delta_\phi(d)$ of the measured AoA from the geometric angle of incidence (AoI) into the Rx. This is measured simply as the difference (in degrees) of the AoA and AoI for each Tx-Rx link with distance $d$. A ``canonical'' LOS scenario such as \textbf{INT}-W-N should have very low values of $\Delta_\phi(d)$, while a ``canonical'' NLOS scenario like \textbf{INT}-W-S should have relatively large values of $\Delta_\phi(d)$.}

\addedMK{Fig.~\ref{fig:aoi} presents a CDF of $\Delta_\phi(d)$ as measured for several scenarios at \textbf{INT}. Fig.~\ref{fig:aoi}\subref{fig:aoi-intw} shows that \textbf{INT}-W-N has a median $\Delta_\phi(d)$ of 2$^\circ$, whereas \textbf{INT}-W-S has a median $\Delta_\phi(d)$ of 31$^\circ$. This is a numerical interpretation of the high-level angular spread that can be observed in Fig.~\ref{fig:stack_spectra}\subref{fig:stack_NLOS} compared to Fig.~\ref{fig:stack_spectra}\subref{fig:stack_LOS}. Furthermore, a reflective signal propagation mechanism is considered by defining the AoI relative to the building across the street halfway between the Tx and Rx, rather than relative to the Tx itself. Defining the AoI in this way, shown by Fig.~\ref{fig:aoi}\subref{fig:aoi-intws-pas}, produces a CDF with considerably lower median $\Delta_\phi(d)$ (3$^\circ$), suggesting that for the \textbf{INT}-W-S scenario, a majority of Tx-Rx links achieve maximum received power via a reflection from the building opposite the Rx. In this specific scenario, the building facade opposite the Rx, seen in Fig.~\ref{fig:aoi}\subref{fig:aoi-intws-opposite} is predominantly made of brick, with a reflection coefficient of $\Gamma \approx 0.4$~\cite{vargas2018measurements,landron1993insitu}.}

\addedMK{Fig.~\ref{fig:aoi}\subref{fig:aoi-intn} shows equivalent results for \textbf{INT}-N-E and \textbf{INT}-N-W. As with the LOS \textbf{INT}-W-N case, the predominantly LOS \textbf{INT}-N-E has a low median $\Delta_\phi(d)$ of 2$^\circ$. \mbox{\textbf{INT}-N-W} is predominantly NLOS, with blockage caused by the building corner opposite the Rx. In this case, a diffraction-based propagation mechanism can be considered~\cite{du2021directional}, implying at a high level that the signal arrives from the direction of the building corner itself. By setting the AoI to be relative to this location, the median $\Delta_\phi(d)$ for \textbf{INT}-N-W is reduced from 9$^\circ$ to 3$^\circ$, a significant reduction given the small change in AoI from Tx to Rx that occurs over this measurement scenario. This result suggests that the signal propagates via diffraction at the building corner, rather than a reflection from the opposite side of the street.}

\begin{figure}[t]
\centering
\includegraphics[width=1.0\columnwidth]{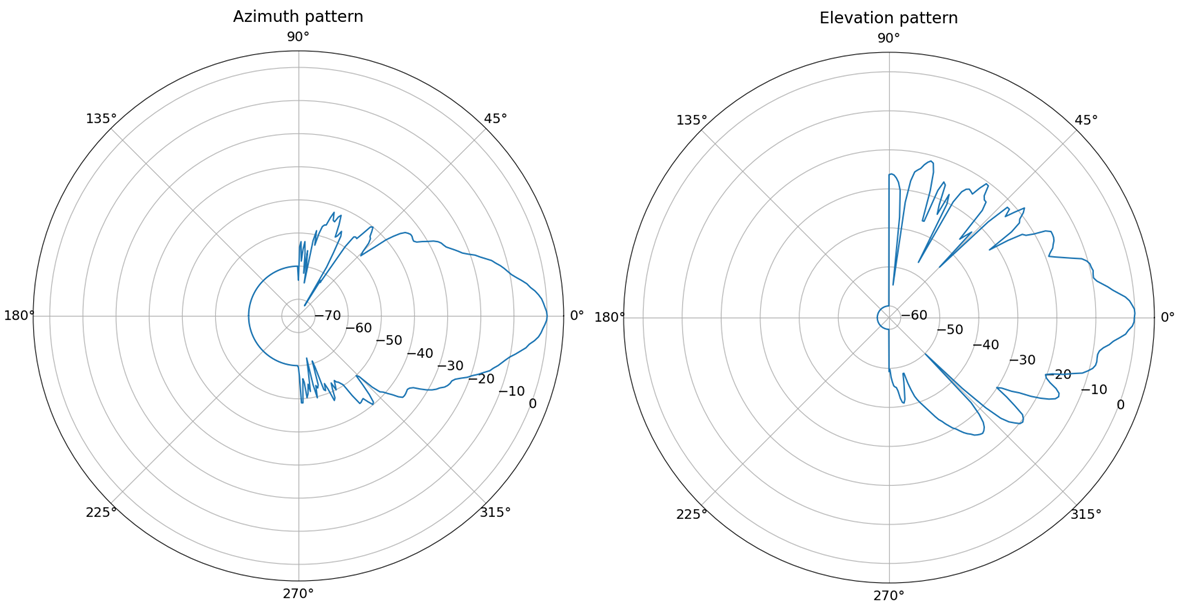}
\vspace{-\baselineskip}
\caption{Horn antenna radiation patterns (normalized)}
\label{fig:horn_antenna}
\vspace{-\baselineskip}
\end{figure}

\subsection{Azimuth Beamforming Gain}
\addedMK{The PAS are helpful not only for understanding specific propagation mechanisms but also measuring the overall angular spread of the 28\thinspace{GHz} signal as it propagates. This angular spread is intuitively quantified as the effective azimuth beamforming gain (ABG), or $G_\text{az}$, and is calculated with Eq.~\ref{eq:abg} in Section~\ref{ssec:meas-metrics}. As the urban environment necessarily introduces more reflections and other scattering effects compared to the anechoic chamber used to profile the RX antenna, the ABG will always be lower than the nominal azimuthal antenna gain measured within such; 14.5\thinspace{dBi} as described in~\ref{ssec:meas-equipment}. The difference in median ABG and the nominal antenna gain is the median \textit{beamforming gain degradation}. For reference, the anechoic chamber measurement of the antenna is provided in Fig.~\ref{fig:horn_antenna}.}

\begin{figure*}[t]
\centering
\vspace{-1\baselineskip}
\subfloat[]{
\includegraphics[width=0.45\linewidth]{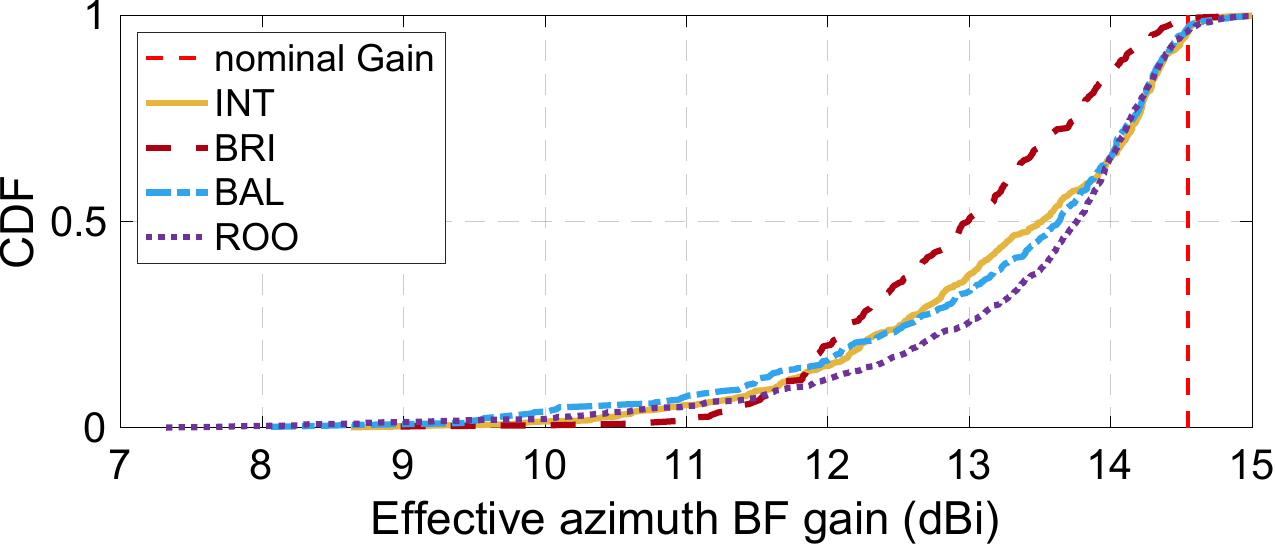}
\label{fig:abg_all_loc}}
\hspace{10pt}
\subfloat[]{
\includegraphics[width=0.45\linewidth]{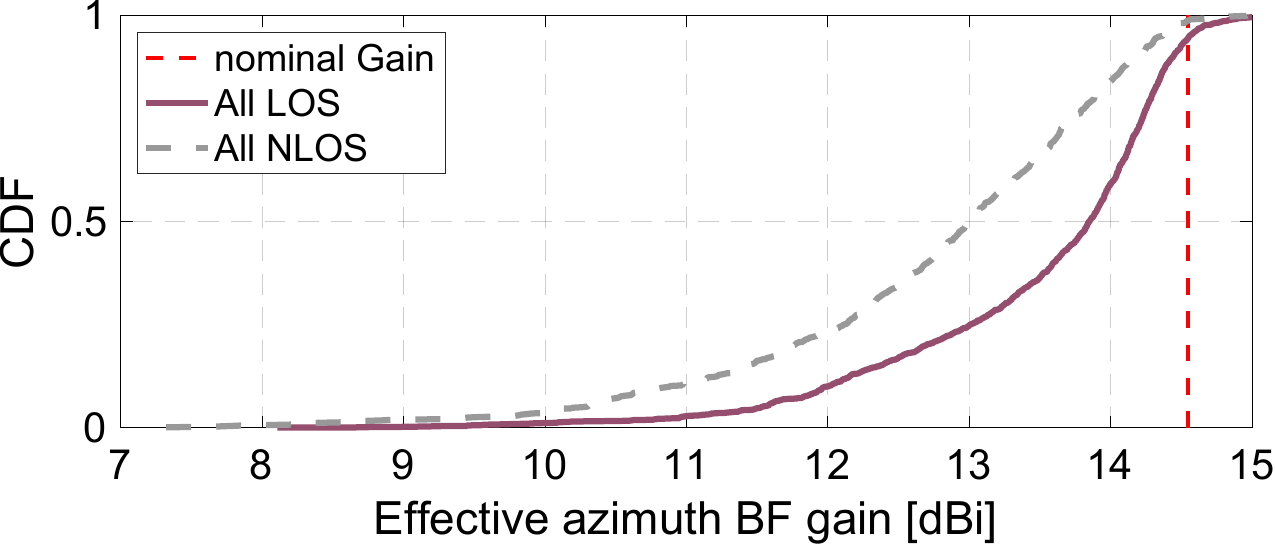}
\label{fig:abg_all}}
\vspace{-0\baselineskip}
\caption{\addedMK{ABG for (a) each Rx location; (b) all measurements separated according to empirical LOS or NLOS.}}
\vspace{-1\baselineskip}
\label{fig:abg}
\end{figure*}

\addedMK{The results in Section~\ref{sec:validation} and Fig.~\ref{fig:placement} include measurements of the ABG. In particular, we noted that NLOS scenarios, such as those in Fig.~\ref{fig:placement}\subref{fig:txrxswap-compare-bfg} and Fig.~\ref{fig:placement}\subref{fig:nextstreet-compare-bfg}, typically have a lower measured ABG compared to LOS scenarios. Furthermore, Fig.~\ref{fig:placement}\subref{fig:txrxswap-compare-bfg} shows that locating the RX within the urban canyon emulating a UE produces 2.5--3.0\thinspace{dB} lower ABG compared to locating the RX on a building emulating a BS.}

\addedMK{We now evaluate the ABG by grouping the measurement scenarios by Rx location and by LOS or NLOS. These results are given as CDFs in Fig.~\ref{fig:abg}. Fig.~\ref{fig:abg}~\subref{fig:abg_all_loc} shows that \textbf{INT}, \textbf{BAL}, and \textbf{ROO} have similar ABG distributions above the 60\textsuperscript{th} percentile, and their median ABG differs by at most 0.5\thinspace{dB}. The \textbf{BRI} Rx location has a slightly lower median compared to the other three locations; this is likely due to the presence of two directions with high received power that was commonly observed at this location. The Rx located above the middle of the street produces a highly symmetric propagation environment, leading to this effect.}

\addedMK{Fig.~\ref{fig:abg}~\subref{fig:abg_all} shows the ABG for all scenarios in Table~\ref{tab:meas-summary} separated by empirical LOS and NLOS (i.e., equivalent to the separation of the data in Fig.~\ref{fig:all-pg}). The median ABG as measured for the NLOS scenarios is 13\thinspace{dBi} while for LOS scenarios is 14\thinspace{dBi}. This shows that the measurements taken in  empirical NLOS scenarios suffer three times higher beamforming gain degradation compared to the LOS scenarios; around 1.5\thinspace{dB} compared to 0.5\thinspace{dB}.} 

\addedMK{These results suggest that placing a BS antenna within a realistic urban environment leads to a 0.5--1.5\thinspace{dB} beamforming gain degradation depending on the LOS or NLOS nature of the UE location on the street. According to Fig.~\ref{fig:placement}\subref{fig:txrxswap-compare-bfg}, this degradation is even more severe for the street UE. There are two resulting consequences. First, this reduces the link budget benefit brought by highly directional antennas; second is a reduced benefit of precise beam alignment algorithms between a BS and UE. Consequentially, this implies that a fine-grained beam searching algorithm would suffer from diminishing returns; instead, a beam search algorithm with lower overhead can be used, such as one that searches over a predefined codebook of beam directions and beam widths.}

\section{SNR Coverage}
\label{sec:coverage}
%

\begin{figure*}[t]
\centering
\vspace{-1\baselineskip}
\subfloat[]{
\includegraphics[height=1.7in]{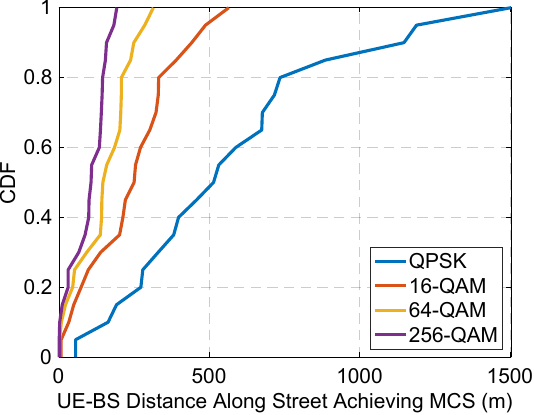}
\label{fig:snr_loc}}
\hspace{5pt}
\subfloat[]{
\includegraphics[height=1.7in]{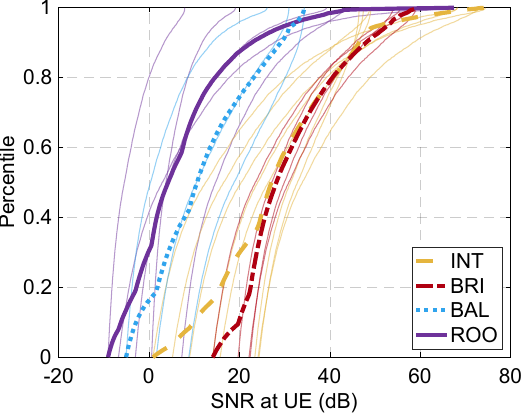}
\label{fig:snr_ue_singleended}}
\hspace{5pt}
\subfloat[]{
\includegraphics[height=1.7in]{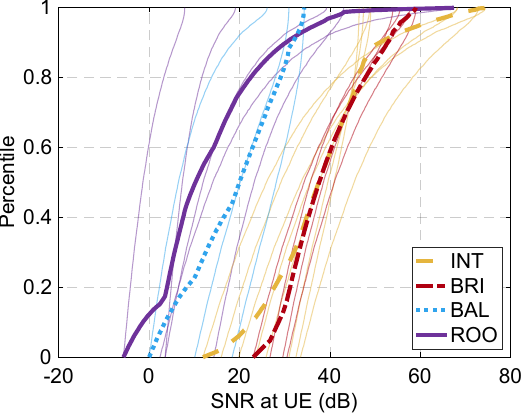}
\label{fig:snr_ue_symmetric}}
\vspace{-0\baselineskip}
\caption{\addedMK{SNR coverage analysis for the locations and scenarios measured. (a) CDF across all measured scenarios of the distances for which the 5G NR FR2 modulation schemes are supported; CDF of 90\textsuperscript{th} percentile SNR across 10,000 simulated UEs for each scenario and location with (b) a BS at one end of the street only, and (c) a BS at each end of the same side of the street.}}
\vspace{-1\baselineskip}
\label{fig:snr_coverage}
\end{figure*}

\addedMK{We now utilize the path gain models described in Section~\ref{sec:results} and summarized in Table~\ref{tab:meas-summary} in a preliminary analysis of the distribution of SNRs achieved by UEs located on street sidewalks, as well as the relationship between the BS-UE distance and the feasible modulation schemes available to the UE. This analysis can provide insight into outdoor-to-outdoor coverage at 28\thinspace{GHz} in dense urban environments, including achievable data rates.}

\subsection{Coverage Analysis}
\addedMK{For the analysis in this section, we assume a set of system parameters. The BS has a TX power of $P_{\textrm{TX}} = +$28\thinspace{dBm} and TX antenna gain of 23\thinspace{dBi}. The resulting total EIRP of $+$51\thinspace{dBm} is comparable to that of typical 28\thinspace{GHz} phased array antenna modules~\cite{sadhu201728} suitable for deployment in an urban BS. The UE is assumed to have a nominal Rx gain of $G_{\textrm{RX}} = $ 3.0\thinspace{dBi} from an omnidirectional antenna and a noise figure of 10\thinspace{dB}.}

\addedMK{For each sidewalk, the effective TX gain, denoted by $G_{\textrm{TX}}$, is obtained by subtracting the median azimuth gain degradation indicated in Table~\ref{tab:meas-summary} from the nominal antenna gain of 23\thinspace{dBi}. Given a {50}\thinspace{MHz} 5G NR FR2 channel, the resulting RX noise floor can be calculated as $P_{\textrm{N}} = -174 + 10\log_{10}(50 \times 10^{6}) + 10 = -87$\thinspace{dBm}. 50\thinspace{MHz} represents the lower end of the available 5G NR FR2 downlink bandwidths, reflecting a likely link condition within the dense urban environment described in Section~\ref{ssec:meas-environment}.}

\addedMK{Given this noise floor, the SNR of a given link at a distance $d$ is then computed as:}
\begin{align}
\text{SNR}(d) & = P_{\textrm{TX}} + G_{\textrm{TX}} + G_{\textrm{path}}(d) + G_{\textrm{RX}} - P_{\textrm{N}},
\end{align}
\noindent \addedMK{where the terms $G_\text{TX}$ and $G_\text{path}(d)$ are street-specific parameters, and the remaining terms are constant. Note that this expression assumes a clear channel, with the only noise present within $P_N$. In practice, noise may be caused by transmissions in adjacent FR2 channels, and we investigate the implications of this from the spatial point of view in Section~\ref{sec:scm}.}

\addedMK{Consideration of the SNR achievable by UEs located on the street sidewalk is helpful to understand what kind of modulation and coding scheme (MCS), and hence data rate, would be achievable for a given link. The BS and UE will negotiate the MCS via the channel state information and the channel quality indicator, which will contain the SNR and similar metrics. We analyze the SNR across the measurement scenarios in Table~\ref{tab:meas-summary} by analyzing at what distance each MCS is supported up to by considering the SNR requirement for each MCS~\cite{3gpp256qamstudy,3gppperformance,peralta20185g}. In this analysis, we disregard the ROO-B-N, ROO-B-S, and ROO-E-S scenarios as they have a positive slope for $G_\text{path}(d)$, a result arising from heavy blockage at short distances. Other propagation mechanisms dominate these streets as measured from \textbf{ROO} rather than those assumed by the 3GPP UMi model~\cite{adhikari2025aroundcorner}, and as such we disregard them in this analysis.}

\subsection{Results}
\addedMK{The results of this coverage analysis are shown in Fig.~\ref{fig:snr_coverage}\subref{fig:snr_loc}. These results assume the presence of a singular BS that has correctly aligned its antenna with that of the UE. Fig.~\ref{fig:snr_coverage}\subref{fig:snr_loc} shows that this singular BS, in the median case, can can support 256-QAM, 64-QAM, 16-QAM, and QPSK up to 100\thinspace{m}, 150\thinspace{m}, 250\thinspace{m} and 525\thinspace{m}, respectively. We note that these results are computed for the 10\textsuperscript{th} percentile user, as such, these are the distances for which 90\% of users at that distance will be supported by the given MCS.}

\addedMK{The results in Fig.~\ref{fig:snr_coverage}\subref{fig:snr_loc} imply that 256-QAM and 64-QAM are unsupported for the 10\textsuperscript{th} percentile user for some of the measured scenarios where only a single BS is assumed. However, in practice, BS availability may be more widespread. It is known that a dense BS deployment on a large number of city block corners can help coverage~\cite{du2021directional}, and mid-block deployments can help spatial diversity techniques~\cite{kohli2024outdoor}. We now investigate how a second base station placed at the far side of the street (in relation to the RX location during the measurements) can impact the SNR coverage for UEs. For this purely mathematical analysis, we simulate the 10\textsuperscript{th} percentile SNR for 10,000 UEs randomly located on the street sidewalk corresponding to each scenario. The location of each UE, and hence its distance from the BS, is governed by a simple first-order approximation of street pedestrian distribution, whereby a pedestrian is twice as likely to be at the end of the street than they are in the middle, representing the convergence of two sidewalks. We repeat this analysis twice, one for the case of a single BS, shown in Fig.~\ref{fig:snr_coverage}\subref{fig:snr_ue_singleended}, and again when another BS is assumed at the far end, shown in Fig~\ref{fig:snr_coverage}\subref{fig:snr_ue_symmetric}. When a second BS is assumed in the simulation, a UE will utilize the one closest to it. On account of the street canyon symmetry, we assume an identical channel model from the second, hypothetical BS.}

\addedMK{From Fig.~\ref{fig:snr_coverage}\subref{fig:snr_ue_singleended}, we see that around 50\% of UEs at the \textbf{ROO} location achieve less than 5\thinspace{dB} SNR 10\% of the time. 5\thinspace{dB} SNR is around the threshold for a QPSK data transmission, this suggests that for 50\% of users, there would be an unacceptably high probability that the UE-BS link is infeasible when only one BS is present. However, should a second BS be present at the far end of the measured street, this figure drops to 18\% of UEs. We note that this is still a very conservative analysis. Consider the \textbf{ROO-S-*} and \textbf{ROO-S2-*} scenarios covering more than 1\thinspace{km} sidewalk distance; it is very possible that more than two BSes would be deployed within this distance, especially with newer infrastructure deployments such as lightpoles or small cells~\cite{Link5G}. Lastly, Fig.~\ref{fig:snr_coverage}\subref{fig:snr_ue_symmetric} shows that for the \textbf{INT} and \textbf{BRI} locations, 100\% of UEs achieve better than 10\thinspace{dB} SNR 90\% of the time, suggesting good coverage for users when two BS are used, and ample margin for use of higher MCS and bandwidths above 50\thinspace{MHz}.}

\section{Spectrum Consumption Models}
\label{sec:scm}
\addedCC{In addition to supporting the generation of path gain models, we leveraged the collected measurements to develop a methodology and set of software tools/scripts to generate Spectrum Consumption Models (SCMs) as standardized by the IEEE 1900.5.2 standard \cite{scmstdofficial} for several of the 28GHz operation cases/locations described in section \ref{sec:meas}.}


\begin{figure}[t]
\centering
\includegraphics[width=1.0\columnwidth]{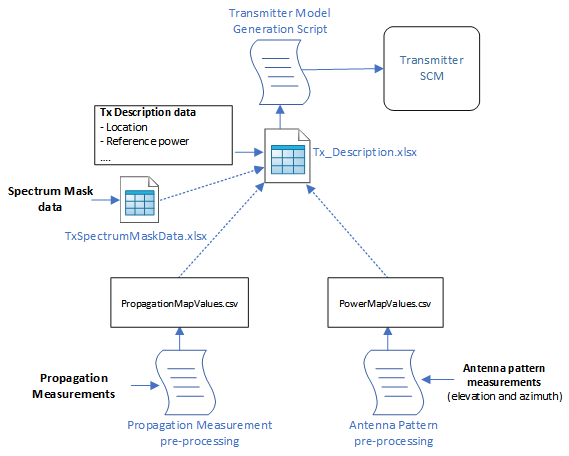}
\vspace{-\baselineskip}
\caption{Process to generate an SCM for a TX}
\label{fig:txscmgen}
\vspace{-\baselineskip}
\end{figure}

\subsection{SCM Generation}
\addedCC{The measured path gain values for several of the locations mentioned in section \ref{sec:meas} were used to populate the information for the “Propagation Map” construct of the mmWave SCMs we created. In SCMs, this construct specifies the rate of RF power attenuation by direction. Thus, in contrast to the traditional fitted line models that are applied to path gains such as those described in section \ref{sec:results}, SCMs can capture directional details on the attenuation of RF power which is highly relevant when using directional antennas and beam forming in mmWave environments.}

\addedCC{The Horn antenna with 24 dBi gain that was used in the measurements was characterized in detail and the Azimuth and Elevation antenna patterns for it, shown in Figure  ~\ref{fig:horn_antenna}, were used to generate a three-dimensional description of the gain of the antenna which was incorporated into the “Power Map” construct of our mmWave SCMs. The high directionality of the antenna and the capability for SCMs to be used to determine compatibility (non-interference) among many RF devices/systems operating in a given area will be useful in the design and study of spectrum sharing mechanisms and wireless network operations for mmWave systems.}

\addedCC{SCMs can be structured in XML or JSON formats. With the high level of detail from the path gain and antenna characterization measurements, the size of the SCMs can grow to several hundred Kilobytes if processed in raw form.  Managing these files is not a big issue but structuring the information contained in them so that efficient searches for specific values (e.g. path gain and antenna gain in a particular direction) can slow down compatibility computations if not properly addressed.}

\addedCC{Figure~\ref{fig:txscmgen} provides a summary of our SCM generation process for the case of a TX device. A GUI-based tool for SCM generation (which is currently being updated) was described in \cite{caicedo2017spectrum} but given the amount of values that need to be entered, our script based process is more manageable since it can process files that contain a large number of measurements.}

\addedCC{In our approach, for simplicity, we use a set of python-based scripts that process Excel spreadsheets and CSV files to facilitate the input and pre-processing of the necessary data/values to generate an SCM. Propagation and antenna pattern measurements are pre-processed to produce a set of files with the values for  the Propagation Map and Power Map constructs of the SCM. A description of the shape of the TX's spectral mask and center frequency for its transmissions is provided via a spreadsheet as input to generate the SCM's spectrum mask construct. Additional information to characterize the location, operation schedule and reference power, among other characteristics, is also provided via a spreadsheet which additionally includes the file paths to all the input files for spectrum mask, propagation map and antenna pattern values. Finally, a Python script reads all the inputs and generates an SCM.}

\subsection{mmWave Spectrum Management with SCMs}
\begin{figure}[t]
    \centering
    \includegraphics[width=8.5cm]{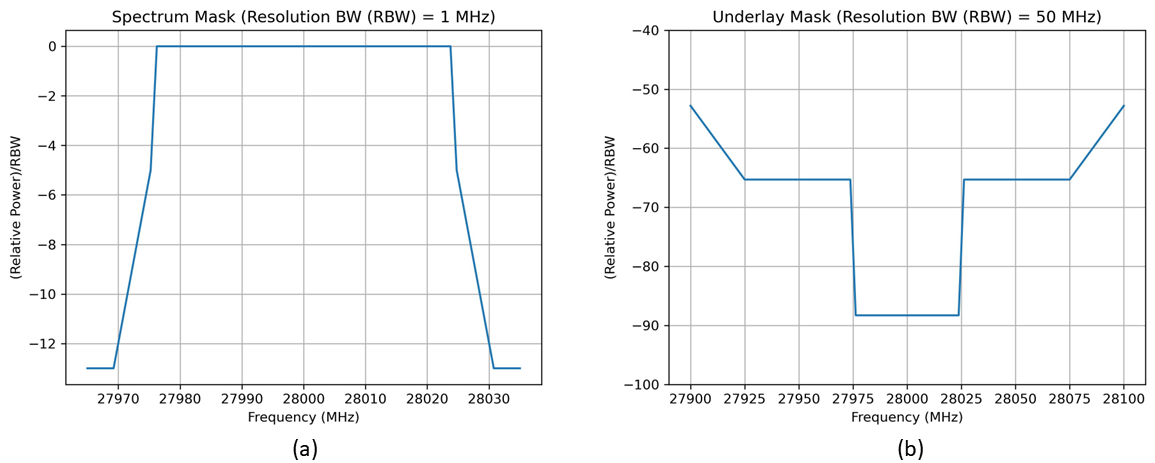}
    \caption{a) FR2 TX spectrum mask b) FR2 UE RX underlay mask }
    \label{fig:scm_masks}
\end{figure}

\addedCC{For the evaluation of the performance and usability of the mmWave SCMs that we generated, we used a custom Python-based simulator and DSA analysis tool described in our previous work in \cite{scmwcnc2023}. The tool was enhanced to support mmWave SCMs and to handle directional antennas and propagation maps. 
In this sub-section we will present two cases where we will use SCMs to model 5G FR2 operations at 28\thinspace{GHz} that capture the directional horn antenna radiation pattern characteristics, antenna orientation configuration parameters and the directional propagation effects that can be derived from the measurements described in section V in order to show how SCMs can be used in practical spectrum deconfliction and/or spectrum management/planning scenarios. We will consider a 5G FR2 base station TX and a 5G FR2 UE RX with the operational characteristics mentioned in Table~\ref{tab:fr2_table}. The TX's spectrum emission mask is shown in Figure~\ref{fig:scm_masks}a and the UEs RX underlay mask is shown in Figure~\ref{fig:scm_masks}b. They follow the emission and reception limits specified for 5G FR2 operations in the 3GPP technical specifications TS 38.104 and TS 38.101 respectively.}

\begin{table}
    \centering
    \begin{tabular}{|c|c|} \hline 
         \multicolumn{2}{|c|}{\textbf{FR2 Base station}}\\ \hline 
         Antenna type& Horn Antenna\\ \hline 
         Antenna gain& 24 dBi\\ \hline 
         Antenna pattern& Directional with  10$^{\circ}$ HPBW. See figure~\ref{fig:horn_antenna}\\ \hline 
         Base station type& Local Area base station\\ \hline 
         Transmit power& 33 dBm\\ \hline
         Channel bandwidth& 50 MHz\\ \hline
         \multicolumn{2}{|c|}{\textbf{FR2 UE}}\\ \hline 
         UE type& Handheld UE - Power class 3\\ \hline 
         Antenna type& Omnidirectional\\ \hline 
         Antenna gain& 3 dBi\\ \hline
         Channel bandwidth& 50 MHz\\ \hline
    \end{tabular}
    \caption{Operational characteristics for modeled FR2 systems}
    \label{tab:fr2_table}
\end{table}

\begin{figure*}[t]
\vspace{-2\baselineskip}
\centering
\subfloat{
\includegraphics[height=3.5in]{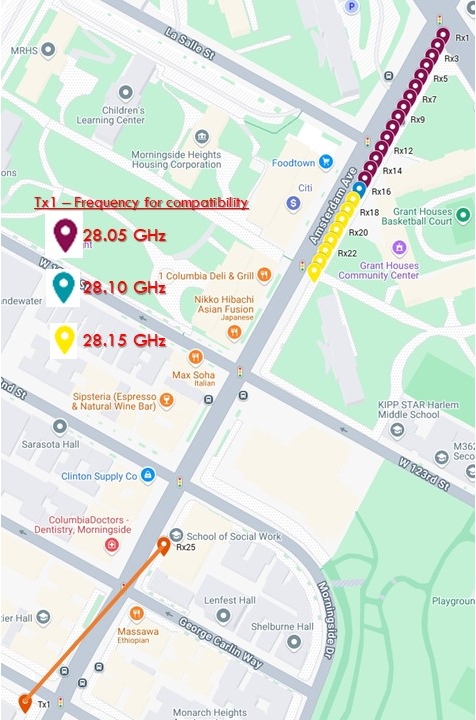}
\label{fig:intNE_case}
}
\hspace{3pt}
\subfloat{
\includegraphics[height=3.5in]{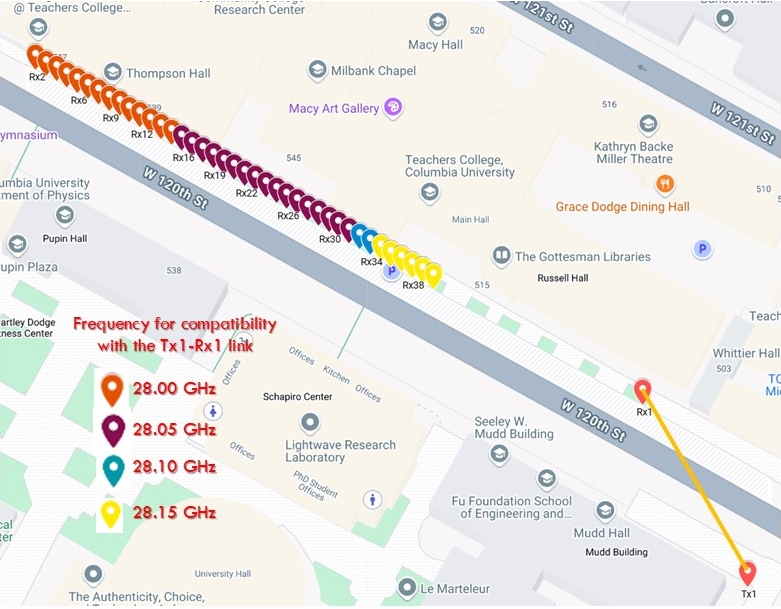}
\label{fig:IntWN_case}
}
\caption{Maps of the mmWave FR2 scenarios. (Left) INT-N-E (Right) INT-W-N}
\label{fig:fr2_cases}
\vspace{-\baselineskip}
\end{figure*}

\addedCC{In our first case, shown in Figure~\ref{fig:fr2_cases}, we use the measurements for the INT-N-E location and consider the scenario where there may be UEs operating at 28.0\thinspace{GHz} in any of the locations labeled RX1 to RX24. In an equivalent manner we could consider that there is a UE moving between locations RX1 to RX 24 and operating at 28.0\thinspace{GHz}. If we wanted an FR2 base station to establish a link with a UE in location RX25, the question we want to answer is what should the transmit frequency of the base station be in order to not interfere with any UE located between RX1 and RX24 that is already operating at 28.0\thinspace{GHz}. The spectrum use compatibility analysis carried out in our DSA analysis tool with the SCMs for this case, indicates that the TX1 to RX25 link should operate in a 50MHz channel with a center frequency other than 28.0\thinspace{GHz} and the closest center frequency it can operate at is 28.05\thinspace{GHz}. At that center frequency the operations of the link will not interfere with a UE RX operating in locations RX1 to RX15 but other center frequency values are needed to avoid interference with a UE RX at locations RX16 to RX24. Figure~\ref{fig:fr2_cases} shows a summary of our analysis. 

For our second case, we use the measurements for the INT-W-N location. In this case, we assume that we have a link already operating at 28.0\thinspace{GHz} between TX1 and RX1 and want to find out the channel center frequency closest to 28.0\thinspace{GHz} at which UE RXs located between locations RX2 and RX40 can operate so that their operation is not prevented by the interference from the TX1 to RX1 link. For this case, our SCM based spectrum use analysis shows that a UE in locations RX2 to RX15 can operate its 50 MHz channel at a center frequency of 28.0\thinspace{GHz}. Thus, in these locations the channel centered at 28.0\thinspace{GHz} can be re-used/shared with the TX1-RX1 link. There is enough angular separation between the orientation of the TX1-RX1 link antennas towards the RX2 to RX15 locations, which along with the directional characteristics of TX1's antenna and of the propagation environment allow as an aggregate for the reuse of the 28.0\thinspace{GHz} channel.  Figure~\ref{fig:fr2_cases}(right side) shows the center frequencies for compatibility with the TX1-RX1 link for other locations.}

\subsection{Future Directions for the Use of mmWave SCMs}
\addedCC{Given the diverse set of locations in which measurements were taken, we will explore approaches for deriving the propagation characteristics for a specific street sidewalk based on the measurements taken on the opposite sidewalk of the same street. Another objective is to devise methods for reducing the number of measurements needed to characterize the propagation and path loss in an urban canyon so that the SCM for a particular sidewalk can be transferred and adapted to an SCM that also describes the propagation characteristics of a different sidewalk and/or similar street canyon.


We also plan to conduct more experiments with PAAMs and SCMs where beam forming and/or the use of multiple beams are leveraged in the design of efficien spectrum use deconfliction methods for mmWave operations.}


\section{Conclusions}
\label{sec:conclusions}
In this paper, we presented the results of an extensive channel measurement campaign in the COSMOS testbed deployment area. \addedMK{We thoroughly studied 4 sites and obtained measurements of over 3,000 links on 24 sidewalks}, some of which were measured multiple times in different settings. We studied the effects of swapped Tx and Rx locations, Tx heights, and seasonal effects on the path gain and concluded that these effects are insignificant in most scenarios. Moreover, we studied the path gain values, the effective azimuth beamforming gain, and median SNR values that can be achieved on the 13 sidewalks. Overall, the results can inform the deployment of the {28}\thinspace{GHz} PAAMs in the COSMOS testbed. \addedCC{Through the generation of SCMs for the 28 GHz cases covered in this study, the measurements can be used to analyze spectrum sharing scenarios in similar urban canyon environments and to enhance the sharing and re-usability of the measurements with other researchers}. However, there are several directions for future work, including more extensive studies of seasonal effects in tree-sparse urban street canyons, and wideband channel measurements for obtaining more detailed channel parameters (e.g., power delay profile) in the same environment.


\bibliographystyle{IEEEtran}
\bibliography{paper_28ghz.bib, oti_ton.bib}

\end{document}